\documentclass[useAMS,usenatbib]{mn2e}
\usepackage{graphicx}
\usepackage{supertabular}
\usepackage[authoryear]{natbib}

\newcommand{\kms}{km\,s$^{-1}$}
\newcommand{\vs}{$v \sin i$}
\newcommand{\teff}{$T_{\rm eff}$}
\newcommand{\lgg}{$\log\,{g}$}
\newcommand{\bz}{\ensuremath{\langle B_z \rangle}}
\newcommand{\bs}{\ensuremath{\langle B \rangle}}
\def\gtrsim{\mathrel{\hbox{\rlap{\hbox{\lower4pt\hbox{$\sim$}}}\hbox{$>$}}}}
\def\ltsim{\mathrel{\hbox{\rlap{\hbox{\lower4pt\hbox{$\sim$}}}\hbox{$<$}}}}

\title[HD 133880]{An Analysis of the Rapidly Rotating Bp Star HD 133880\thanks{Based in part on observations made with the European Southern Observatory telescopes under ESO programmes 082.D-0061(A), 083.D-0034(A), 	085.D-0296(A), 086.D-0449(A), obtained from the ESO/ST-ECF Science Archive Facility}}
\author[J. D. Bailey et al.]
{J. D. Bailey,$^{1}$\thanks{Based in part on observations obtained at the Canada-France-Hawaii Telescope (CFHT) which is operated by the National Research Council of Canada, the Institut National des Science de l'Univers of the Centre National de la Recherche Scientifique of France, and the University of Hawaii.} J. Grunhut,$^{2}$ M. Shultz,$^{2}$ G. Wade,$^{2}$ J. D. Landstreet,$^{1,3}$ D. Bohlender,$^{4}$ 
\newauthor 
J. Lim,$^{5,6}$ K. Wong,$^{5}$ S. Drake,$^{7}$ J. Linsky,$^{8}$ and the MiMeS Collaboration\\
$^{1}$Physics \& Astronomy Department, The University of Western Ontario, London, Ontario, Canada N6A 3K7\\
$^{2}$Department of Physics, Royal Military College of Canada, Kingston, Ontario, Canada K7K 7B4\\
$^{3}$Armagh Observatory, College Hill, Armagh, Northern Ireland\\
$^{4}$Herzberg Institute of Astrophysics, National Research Council of Canada, 5071 West Saanich Road, Victoria BC, Canada V9E 2E7\\
$^{5}$Department of Physics, University of Hong Kong, Pokfulam Road, Hong Kong\\
$^{6}$Institute of Astronomy \& Astrophysics, Academia Sinica, Taipei 10617, Taiwan\\
$^{7}$USRA \& NASA, Code 662, Goddard Space Flight Center, Greenbelt, Maryland, 20771, USA\\
$^{8}$JILA, Campus Box 440, University of Colorado, Boulder, Colorado, 80309, USA
}
\begin{document}

\date{Accepted . Received }

\pagerange{\pageref{firstpage}--\pageref{lastpage}} \pubyear{2012}

\maketitle

\label{firstpage}

\begin{abstract}
HD~133880 is a rapidly rotating Bp star (\vs\ $\simeq$ 103~\kms) and is host to one of the strongest magnetic fields of any Ap/Bp star.  A member of the Upper Centaurus Lupus association, it is a star with a well-determined age of 16 Myr.  Twelve new spectra, four of which are polarimetric, obtained from the FEROS, ESPaDOnS, and HARPS instruments, provide sufficient material from which to re-evaluate the magnetic field and obtain a first approximation to the atmospheric abundance distributions of He, O, Mg, Si, Ti, Cr, Fe, Ni, Pr, and Nd.

An abundance analysis was carried out using ZEEMAN, a programme which synthesises spectral line profiles for stars with permeating magnetic fields.  The magnetic field structure was characterised by a colinear multipole expansion from the observed variations of the longitudinal and surface fields with rotational phase.  Both magnetic hemispheres are clearly visible during the stellar rotation, and thus a three-ring abundance distribution model encompassing both magnetic poles and magnetic equator with equal spans in co-latitude was adopted.

Using the new magnetic field measurements and optical photometry together with previously published data, we refine the period of HD~133880 to $P = 0.877476 \pm 0.000009$~days.  Our simple axisymmetric magnetic field model is based on a predominantly quadrupolar component that roughly describes the field variations.  Using spectrum synthesis, we derived mean abundances for O, Mg, Si, Ti, Cr, Fe, and Pr.  All elements, except Mg, are overabundant compared to the Sun.  Mg appears to be approximately uniform over the stellar surface while all other elements are more abundant in the negative magnetic hemisphere than in the positive magnetic hemisphere.  In contrast to most Ap/Bp stars which show an underabundance in O, in HD~133880 this element is clearly overabundant compared to the solar abundance ratio.

In studying the H$\alpha$ and Paschen lines in the optical spectra we could not unambiguously detect information about the magnetosphere of HD~133880.  However, radio emission data at both 3 and 6~cm suggests that the magnetospheric plasma is held in rigid rotation with the star by the magnetic field and further supported against collapse by the rapid rotation. Subtle differences in the shapes of the optically thick radio light curves at 3 and 6~cm suggest that the large-scale magnetic field is not fully axisymmetric at large distances from the star. 
 
\end{abstract}


\section{Introduction}
HD 133880 (=HR 5624) is a rapidly rotating late B-type chemically peculiar (Bp) star with the Si~$\lambda$4200 peculiarity.  It exhibits an unusual magnetic field.  The mean line-of-sight magnetic field \bz\ is very strong and observed to vary from about -4 to +2~kG \citep{Landstreet1990}.  Unlike most chemically peculiar stars, HD~133880 has a field that is predominantly quadrupolar as opposed to dipolar.   HD~133880 is a photometric variable with variations on the order of 0.15~mag in the $U$-band, which may be the result of the large magnetic field \citep{waelkens1985}.  The magnetic field serves to create a ``patchy'' distribution of elements on the surface of the star.  This non-uniform abundance distribution, and its associated line blocking and backwarming, may explain the photometric variations \citep{Ryabchikova91}.  Specifically, several elements are distributed non-uniformly over the stellar surface in a non-axisymmetric pattern about the rotation axis.   As the star rotates, the observed magnetic field strength \bz\ varies and over/underabundances of elements are detected.  The most striking anomalous abundances in Bp stars are often found for Cr and rare-earths which can be as much as 10$^{2}$ and 10$^{4-5}$ overabundant compared to the Sun, respectively \citep{Ryabchikova91}.   The resultant variability is explained by the oblique rotator model:  the rotation and magnetic field axes are at angles $i$ and $\beta$ to the line-of-sight and rotation axis, respectively. 

\citet{Landstreet1990} used longitudinal magnetic field measurements obtained from H$\beta$ in conjunction with photometric data reported by \citet{waelkens1985}  to deduce a period of P = 0.877485 $\pm$ 0.00002~days for HD~133880.  \citet{Lim1996} did a preliminary analysis of the rotational modulation of 3~cm and 6~cm radio emission from HD~133880.  The radio emission variation had both broad and narrow peaks that correlated with the maximum extrema of the dipolar, $B_{d}$, and quadrupolar, $B_{q}$, contributions to the magnetic field reported by \citet{Landstreet1990}, respectively.     

\citet{paper2} confirmed HD~133880 as a member of the Upper Cen Lup association and thus it is a star with a well known age of $\log t$ = 7.20 $\pm$ 0.10 (yr).  HD~133880 is a very young star, having completed only about 5\% of its main sequence lifetime \citep{paper2, paper3}.  Their preliminary analysis of many physical characteristics concluded that \teff\ = 12000 $\pm$ 500~K, $\log L/L_{\odot}$ = 2.10 $\pm$ 0.1, and M/M$_{\odot}$ = 3.20 $\pm$ 0.15. (Our analysis (see below) has shown that the \teff\ is likely closer to 13000~K).  Table~1 summarises the physical properties of HD~133880.

There are few hot chemically peculiar stars with broad spectral features for which abundance analyses have been carried out.  Because of the very complex, strong and unusual magnetic field, HD~133880 is a worthwhile candidate to study in detail.  It is also interesting from the fact that it is a star with a well known age.  HD~133880 is one in a large sample of cluster Ap/Bp stars first compiled by \citet{paper1} and \citet{paper2, paper3} to study the time evolution of magnetic fields in Ap stars.  The current study is part of an effort to understand empirically how chemical abundances evolve with time and stellar age in the magnetic chemically peculiar stars.  Recently \citet{Bailey2011} characterised one star in this sample, the Ap star HD~318107 (=NGC 6405 77).  HD~133880 is a second star of great interest from that sample.

This paper will discuss the efforts to characterise the magnetic field and abundance distributions of several elements for HD~133880.  The following section discusses the polarimetric and spectroscopic observations used as well as the deduced longitudinal fields; Sect. 3 discusses improvement of the rotational period ; Sect. 4 describes the determination of physical parameters; Sect. 5 addresses the observed line profile variations; Sect. 6 \& 7 discuss the magnetic field and abundance models; Sect. 8 reports the abundances obtained; Sect. 9 discusses the H$\alpha$ and radio magnetosphere; and Sect. 10 summarises the presented work.

\begin{table}
\centering
\caption{Summary of stellar, wind, magnetic and magnetospheric properties of HD~133880.}
\begin{tabular}{l|rl}
\hline
Spectral type & B8IVp  Si $\lambda$4200   & SIMBAD  \& BSC       \\
$T_{\rm eff}$ (K) & 13000 $\pm$ 600 & This paper\\
log $g$ (cgs) & 4.34 $\pm$ 0.16  & This paper   \\
R$_{\star}$ (R$_\odot$) & 2.01 $\pm$ 0.32 & This paper\\
$v\sin i$ (km\,s$^{-1}$) & 103 $\pm$ 10 & This paper \\
P (d) & 0.877476 $\pm$ 0.000009 & This paper\\
$\log (L_\star/L_\odot)$ & 2.02 $\pm$ 0.10  & This paper\\
$M_{\star}$ ($M_{\odot}$) & 3.20 $\pm$ 0.15 & Landstreet et al. (2007)\\
$\log t$ (Myr) & 7.20 $\pm$ 0.10  & Landstreet et al. (2007)\\
\hline
$B_{\rm d}$ (G) & $-9600 \pm 1000$ & This paper\\
$B_{\rm q}$ (G) & $-23000 \pm 1000$ & This paper\\
$B_{\rm oct}$ (G) & $1900 \pm 1000$ & This paper\\
$i$ ($\degr$) & 55 $\pm$ 10 & This paper\\
$\beta$ ($\degr$) & 78 $\pm$ 10 & This paper\\
\hline
$\eta_*$ & $\sim 10^{7}$ & This paper\\
$W$ & 0.3 & This paper\\
$\tau_{\rm spin}$ & 11 Myr & This paper\\
\hline\hline
\end{tabular}
\label{params}
\end{table}

\section{Observations}

\subsection{Polarised spectra}
High-resolution spectropolarimetric (Stokes $I$ and $V$) observations of HD 133880 were collected with ESPaDOnS at the 3.6m Canada-France-Hawaii Telescope and HARPSpol on the ESO 3.6m telescope. ESPaDOnS and HARPSpol are both high resolution fibre-fed spectropolarimeters. Three ESPaDOnS observations were obtained on August 3, 2010, and on July 14 and 15, 2011 (the latter in the context of the Magnetism in Massive Stars Large Programme (MiMeS)), and one HARPSpol spectrum was obtained on May 3, 2010. Each ESPaDOnS spectropolarimetric sequence consisted of four individual subexposures taken in different retarder configurations, whereas the HARPSpol spectrum results from the combination of four independent spectra, each comprised of four subexposures, obtained over the course of about 9 minutes. The ESPaDOnS measurements cover a spectral range from 370 nm to 1050 nm at a spectral resolution of R$\sim$65~000, while the HARPSpol observation covers 370 nm to 690 nm at a spectral resolution of R$\sim$100~000. 

From each set of four subexposures we derived Stokes $I$ and Stokes $V$ spectra following the double-ratio procedure described by \citet{donati1997}, ensuring in particular that all spurious signatures are removed at first order. Null polarization spectra (labeled $N$) were calculated by combining the four subexposures in such a way that polarisation cancels out, allowing us to verify that no spurious signals are present in the data (see Donati et al. 1997 for more details on the definition of $N$). All ESPaDOnS frames were processed using the CFHT's Upena pipeline, feeding the automated reduction package Libre ESpRIT \citep{donati1997}. The HARPSpol spectrum was reduced using the REDUCE code \citep{reduce2002}.  The details of this reduction can be found in recent papers by \citet{harpspol1-2011} and \citet{harpspol2-2011}.   The peak signal-to-noise ratios (SNRs) per 1.8 \kms\  velocity bin in the reduced ESPaDOnS and HARPSpol spectra range from 449-697.

The log of spectropolarimetric observations is presented in Table~\ref{fulllsd}.

\subsection{Unpolarised spectra}
High resolution spectroscopic observations (Stokes $I$) were obtained from ESPaDOnS at CFHT and ESO's FEROS instrument at the 2.2m telescope located at La Silla.  FEROS is an \'Echelle spectrograph that has a wavelength coverage from 350~nm to 920~nm with a spectral resolution of R$\sim$48000.  FEROS is fed by two fibres (object and sky fibres), which in principle can be used to subtract the contribution of the sky from the object spectrum.  A more detailed description is given by \citet{feros1999}.   The two ESPaDOnS observations were taken on August 4 and 6, 2010, and the six FEROS spectra were collected on February 5, 6 and July 19, 2009 as well as February 7, 8 and 9, 2011.  

The ESPaDOnS data were reduced using Libre-Esprit in a manner similar to that described above.  The FEROS data were reduced using the data-reduction software (DRS) implemented under MIDAS \citep{feros1999}.  The log of unpolarised spectra is presented in Table~\ref{unpolarised}.

\begin{table*}
\caption{Log of ESPaDOnS and HARPSpol observations of HD 133880. Listed are the instrument used, the Heliocentric Julian date of the midpoint of the observation, total exposure time, the peak signal-to-noise ratio per 1.8~\kms\  velocity bin, the phase of the observation (according to Eq. 2), the evaluation of the detection level of a Stokes $V$ Zeeman signature (DD=definite detection, MD=marginal detection, ND=no detection), and the derived longitudinal field and longitudinal field detection significance $z$ from both $V$ and $N$. In no case is any marginal or definite detection obtained in the $N$ profiles. }
\begin{center}
\begin{tabular}{ccccccrrrr}\hline\hline
 & & & & & & \multicolumn{2}{c}{$V$} & \multicolumn{2}{c}{$N$}\\
Instrument & HJD & $t_{\rm exp}$ & SNR & Phase & Detect & $B_\ell\pm \sigma_B$&$z$ & $B_\ell\pm \sigma_B$&$z$ \\
 & & (s) & pix$^{\rm -1}$ & & & (G) & & (G) & \\
\hline   
ESPaDOnS & 2455411.732 & 720 & 629  & 0.640 & DD &$ 2029\pm38$& 53.4 & $ -51\pm31  $ &  1.6 \\
ESPaDOnS & 2455756.795 & 720 & 697 & 0.878 & DD &$ -2103\pm38 $ & 55.3 &   $43\pm23   $ &  1.9 \\
ESPaDOnS & 2455757.877 & 720 & 573 & 0.111 & DD &$ -3671\pm43 $ & 85.4 &   $-23\pm26   $ & 0.88 \\
HARPSpol  & 2455319.726 & 520 & 449 & 0.787 & DD &$ 650\pm62 $& 10.5 & $-15\pm58$ & 0.26   \\
\hline
\label{fulllsd}
\end{tabular}
\end{center}
\end{table*}

\begin{table}
\caption{Log of unpolarised spectra.  Listed are the instrument used, the Heliocentric Julian date, the peak SNR per 1.8~\kms\ velocity bin, the exposure time and the phase computed using the ephemeris of Eq. (2).}
\begin{center}
\begin{tabular}{ccccc}\hline\hline
Instrument & HJD & SNR &   $t_{\rm exp}$ (s) & Phase \\
\hline
ESPaDOnS & 2455412.727  & 180 & 386 & 0.774\\
ESPaDOnS & 2455414.728  & 180 & 394 & 0.054\\
FEROS & 2454867.846  & 130 & 143 & 0.809\\
FEROS & 2454868.818  & 130 & 103 & 0.917\\
FEROS & 2455031.608  & 250 & 122 & 0.438 \\
FEROS & 2455599.820  & 240 & 141 & 0.990\\
FEROS & 2455600.828  & 240 & 151 & 0.139\\
FEROS & 2455601.868  & 300 & 121 & 0.325\\\hline\hline
\label{unpolarised}
\end{tabular}
\end{center}
\end{table}

\subsection{Longitudinal Magnetic Field Measurements}

The circular polarisation observations were analyzed using the multiline analysis technique Least Squares Deconvolution (LSD; Donati et al. 1997). LSD combines the information from essentially all metallic and He lines in the spectrum by means of the assumption that the spectrum can be reproduced by the convolution of a single ``mean" line profile (the LSD profile) with an underlying spectrum of unbroadened atomic lines of specified line depth, Land\'e factor, and wavelength (the line mask, as described by Wade et al., 2000). This process allows the computation of single averaged Stokes \textit{I} and \textit{V} profiles with much higher signal-to-noise ratios than those of the individual lines, dramatically improving the detectability of Zeeman signatures due to stellar magnetic fields. 

The atomic data were taken from the Vienna Atomic Line Database (VALD) \citep{vald3, vald2, vald4, vald1}.  The VALD line list used in this analysis is characterised by the stellar effective temperature $T_{\rm eff}$ and surface gravity $\log g$ reported in Table 1, abundances as derived from the detailed abundance analysis (see Sect. 8), and an unbroadened line depth threshold equal to 1\% of the continuum intensity $I_{c}$, yielding approximately 10000 lines in the ESPaDOnS spectral window and 8000 lines in the HARPSpol window. The line depths were then interactively adjusted so that the line depths predicted by the LSD convolution model matched the observed depth of the spectral absorption lines as well as possible, thus improving the fit between the spectrum and the LSD model. Because Stokes $V$ profiles are detected in many individual spectral lines of HD 133880, we verified (by examining simultaneously the Stokes $V$ and $I$ profiles) that this adjustment procedure also resulted in a better agreement between the observed and computed Stokes $V$ spectrum.

The mean longitudinal magnetic field was measured by computing the first-order moment of the Stokes \textit{V} LSD profile within the line according to the expression: 

\begin{equation}
\bz\ = -2.14\times 10^{11}\frac{\int \! vV(v)\mathrm{d}v}{\lambda g_{\rm{eff}}c\int \left[I_{\rm{c}}-I(v)\right] \mathrm{d}v},
\end{equation}

\noindent where \bz\ is in G, $g_{\rm{eff}}$ is the effective Land\'e factor, $\lambda$ is the mean wavelength of the LSD line in \AA, and \textit{v} is the velocity within the profile measured relative to the centre of gravity (Mathys et al., 1989; Donati et al. 1997; Wade et al., 2000). The uncertainties associated with \bz\ were determined by propagating the formal (photon statistical) uncertainties of each pixel through Eq. (1). The integration ranges employed in the evaluation of Eq. (1) associated with each Stokes profile were selected individually through visual inspection so as to include the entire span of the (highly variable) Stokes \textit{I} profile.  

\subsection{ATCA radio continuum measurements}

HD 133880 was observed with the Australia Telescope Compact Array (ATCA) at 6 cm and 3.5 cm wavelength simultaneously on February 12, 14 and 16, 1995.  On each day, the observation spanned $\sim$10 hours, during which time we observed HD 133880 continuously apart from short ($\sim$3 mins) scans of a secondary calibrator at regular ($\sim$20 mins) intervals.  In this way, we were able to attain full coverage in rotation phase for the star, with repeated coverage (on two separate days) for just over half a rotation phase to check where any observed variations were repeatable (as indeed they were, as shown in Fig. 2 of \citet{Lim1996}).  The primary (absolute flux density) calibrator used was PKS1934-638, and the secondary (complex gain) calibrator PKS1458-391.  We reduced the data in the standard fashion using MIRIAD, being careful to omit data with poor phase coherence (typically at the beginning and end of each day when the source was low in the sky). 

\section{Rotation period}
A period of $0.87746\pm 0.00001$~d was reported by \citet{waelkens1985} based on Geneva photometric observations. Landstreet (1990) recommended an adjustment of the period to $0.877485\pm 0.000020$~d to bring the magnetic measurements reported by \citet{BL1975} into agreement with those that he had acquired in 1987 and 1988.

When we phase the accumulated magnetic measurements of HD 133880 (i.e. the three measurements of \citet{BL1975}, the 12 measurements of \citet{Landstreet1990}, and the 4 new measurements presented here) according to Landstreet's (1990) ephemeris, we observe a systematic offset of $\sim 0.1$ cycles between the new measurements and those published previously. Such an offset could be interpreted as the consequence of a small error in the assumed period. However, given the presence of important abundance non-uniformities on the surface of HD 133880, it is likely that the shape, and possibly the amplitude of the longitudinal field variation derived from metallic lines (i.e. our new LSD measurements) and those derived from hydrogen lines (i.e. the previously published measurements) may differ. Given the sparse phase coverage of our new measurements, it is therefore ambiguous whether this apparent offset is truly a consequence of a period error.

In an attempt to resolve this uncertainty, we examined the photometric measurements obtained by the Hipparcos mission (ESA, 1997). One hundred $H_{\rm p}$-band measurements of HD 133880 are reported in the 1997 edition of the catalogue.\footnote{{\tt vizier.u-strasbg.fr/viz-bin/VizieR-S?HIP\%2074066)}} The periodogram of those measurements in the range 0.5-1.5 days shows a clear, unique peak at $0.877479\pm 0.000030$~d (The uncertainty on the period, which corresponds to 1$\sigma$ confidence, is computed assuming Gaussian statistics by calculating the variation of the reduced $\chi^2$ of a sinusoidal fit to the phased measurements in the vicinity of the best-fit period (e.g. \citet{Press1986}). A similar analysis applied to Waelkens' (1985) 37 $V$-band measurements yields $0.877459^{+0.0001}_{-0.00006}$~d. Both of these periods are consistent within the ($1\sigma$) error bars. To improve the precision of the period, we combined the two datasets. First, we used the transformation reported by \citet{harm1998} to convert the $H_{\rm p}$ measurements to Johnson $V$-band.  We then performed the same period analysis on the combined data set (spanning nearly 15 years), obtaining a period of $0.8774731\pm 0.0000060$~d.

Therefore the photometric data are indicative of a period of 0.8774731~d, intermediate between those of \citet{waelkens1985} and \citet{Landstreet1990}.

We note that if we phase the magnetic data with the above photometric period, a slightly longer period of 0.877476~d is necessary to bring the new magnetic measurements into better agreement with the older measurements. Finally, this period also removes the phase offset between the observed radio flux variation noted by \citet{Lim1996} and the magnetic extrema.

Based on these results, we adopt the following ephemeris for HD 133880:

\begin{equation}
{\rm JD}_{H\alpha^{\rm max}}= 2445472.000(10)+0.877476(9)\cdot E,
\label{ephemeris} 
\end{equation}
 
 \noindent where we have adopted a zero point corresponding to minimum photometric brightness (which occurs simultaneously with the minimum of the longitudinal magnetic field).  We determined this (3$\sigma$) uncertainty by examining the $\chi^{2}$ of the fit to the photometric data about this best-fit value.

\begin{figure}
\begin{center}
\includegraphics[angle=-90,width=0.5\textwidth]{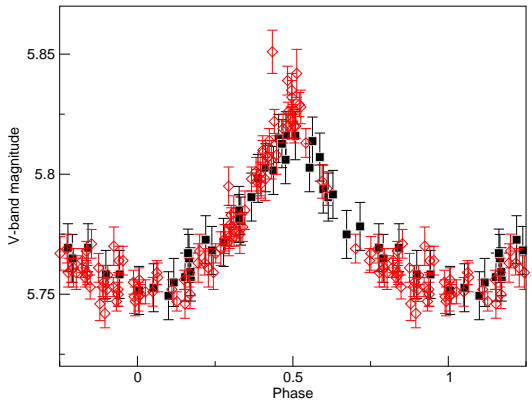}
 \caption{Phased $V$-band (\citet{waelkens1985}; filled squares) and $H_{\rm p}$-band (ESA 1997; open diamonds) photometry. The Hipparcos photometry has been scaled to the Johnson $V$ system using the transformation of \citet{harm1998}.}
\label{phase}
\end{center}
\end{figure}

\section{Determination of Physical Parameters}
\subsection{Effective Temperature}
The enhanced metal abundances and the effects of the magnetic field have to be taken into account when determining the effective temperatures of Ap stars \citep{apteff,hauck-kunzli1996}.  As a result, effective temperatures determined using Geneva and/or Str$\ddot{\rm o}$mgren $uvby\beta$ photometry using normal star calibrations must be corrected to the Ap stars temperature scale.  However, \citet{hauck-north93} suggest that stars that are either He-weak or He-strong do not require that their temperatures be corrected.   

HD~133880 has available both Geneva and Str$\ddot{\rm o}$mgren $uvby\beta$ photometry for the determination of effective temperature.  We have used both in this analysis.  For the Geneva photometry we have used the FORTRAN programme described by \citet{geneva}, and discussed by \citet{paper2}, to obtain an estimate of the effective temperature.  For the Str$\ddot{\rm o}$mgren $uvby\beta$ photometry we used the FORTRAN code ``UVBYBETANEW-ap'' which is a modified version of ``UVBYBETANEW'' \citep{NSW,MD} that corrects the effective temperature to the Ap temperature scale.  However, in this analysis no correction was applied to the temperature because HD~133880 is classified as a He-weak star \citep{hauck-north93}, which is confirmed in this paper (see Sect. 8).

We found \teff\ = 13300~K and 12700~K using $uvby\beta$ and Geneva photometry respectively.  The scatter of these measurements about their mean is of the order of 300~K.   Given the intrinsic uncertainties in determining the effective temperature for a magnetic Ap star in addition to the global uncertainties in the applied methods we optimistically adopt an uncertainty of approximately $\pm 600$~K.  Both the Geneva and $uvby\beta$ photometries are rotationally averaged values and their variations are included within the quoted uncertainties. We use the average of the computed temperatures, adopting an effective temperature of \teff\ = $13000 \pm 600$~K for HD 133880.  If the Ap correction to the temperature is applied, a value of \teff\ = 12440~K is found, which is within the the uncertainty of our adopted value. 

\citet{netopil2008} performed an analysis on the temperature calibration of chemically peculiar stars.  Included in their analysis was HD~133880.  Using photometry, their calibrated temperature corrections suggest that the \teff\ is 11930 $\pm$ 210~K.  This agrees, within uncertainties, to our derived \teff\ if we corrected our derived temperature to the Ap temperature scale.  In our analysis, we favour the higher \teff\ based on work by \citet{hauck-north93} and the better quality fits to H$\beta$ using the model fits of \citet{kurucz1979} with the higher of these two temperatures.  We do, however, compare the derived abundances of all elements using the \teff\ reported by \citet{netopil2008} of 12000~K to our adopted value of 13000~K (see Sect. 8).  

\subsection{Luminosity}
In addition to \teff\ and the $V$ magnitude of HD~133880 (see Table~\ref{params}), we also require a bolometric correction in order to calculate the stellar luminosity.  \citet{paper2} derived a set of bolometric corrections as a function of \teff\ that can be applied to Ap stars.  Applying this correction (see Equation 1 in \citet{paper2}) results in $\log\ L/L_{\odot} = 2.02 \pm 0.10$.  Here, the uncertainty was estimated in the same manner as described in Sect. 1 and by \citet{paper2}.   Note that we used E(B - V) = 0.00 to compute the luminosity, which is confirmed by looking at the colour-colour diagram of (U - B) versus (B - V) 

\subsection{Other Parameters}
Based on values of \teff\ and luminosity determined in Sect. 4.1 and 4.2, the stellar radius is found to be $R = 2.01 \pm 0.32~R_{\odot}$.  \citet{paper2} report a mass for HD~133880 of $M = 3.20 \pm 0.15~M_{\odot}$ (consistent with the values of \teff\ and luminosity in this paper), which can be used with the derived radius to estimate the surface gravity, \lgg.  We find \lgg\ = $4.34 \pm 0.16$, consistent with young age.  

From our adopted period of P = $0.877476\pm 0.000009$~days (Sect. 3) and $v \sin i = 103 \pm 10$~kms$^{-1}$ (see below) we can estimate the inclination angle of the rotation axis to the line of sight, $i$, using the equation
\begin{equation}
\sin i = \frac{(v \sin i)P}{50.6R},
\end{equation}
where $v \sin i$ is in $\rm kms^{-1}$ and R in solar radii, to obtain a value of $i = 63^{\circ} \pm 18^{\circ}$.  All the physical parameters derived are summarised in Table~\ref{params} and are in agreement with those reported by \citet{paper2}.

\section{Line profile variations}
To characterise the line profile variability of HD~133880 we measured the equivalent width variations for a number of spectral lines. The spectral lines were re-normalised to the surrounding continuum regions and the equivalent widths were computed by numerically integrating over the line profile. A single uncertainty value was estimated for each pixel from the RMS scatter in the continuum regions surrounding the line profile and the resulting uncertainties in the equivalent width measurements were found by adding the pixel uncertainties in quadrature.

\begin{figure}
\centering
\includegraphics[width=3.2in]{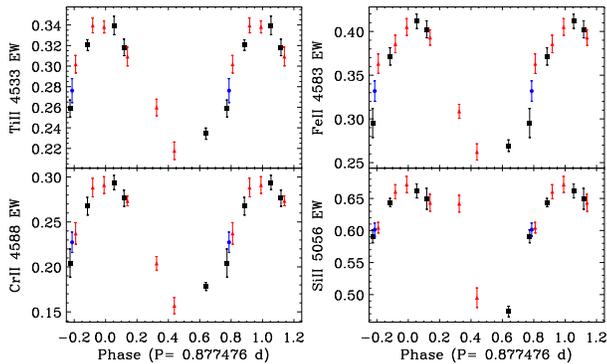}
\caption{Phased equivalent widths measurements for the indicated spectral lines measured from the ESPaDOnS (black squares), FEROS (red triangles) and HARPSpol (blue circles) datasets.}
\label{ew_plots}
\end{figure}

In Fig.~\ref{ew_plots} we show equivalent width measurements phased to the ephemeris given in Eq.~(\ref{ephemeris}) for select lines of titanium, iron, chromium and silicon. The equivalent width curves are all similar and show strong, approximately sinusoidal variations consistent with the rotation period. For each element, the equivalent width reaches a minimum at about phase 0.5. The largest variation is seen in the Cr\,{\sc ii} 4588\,\AA\ line with the equivalent width measurements almost doubling in value. Similar equivalent width changes are found in the Ti\,{\sc ii} 4533\,\AA\ and Fe\,{\sc ii} 4583\,\AA\ lines, but a much smaller amplitude of the variation is measured in the Si\,{\sc ii} 5056\,\AA\ doublet. Clear variations are also seen in other spectral lines of these elements.

\begin{figure}
\centering
\includegraphics[width=1.6in]{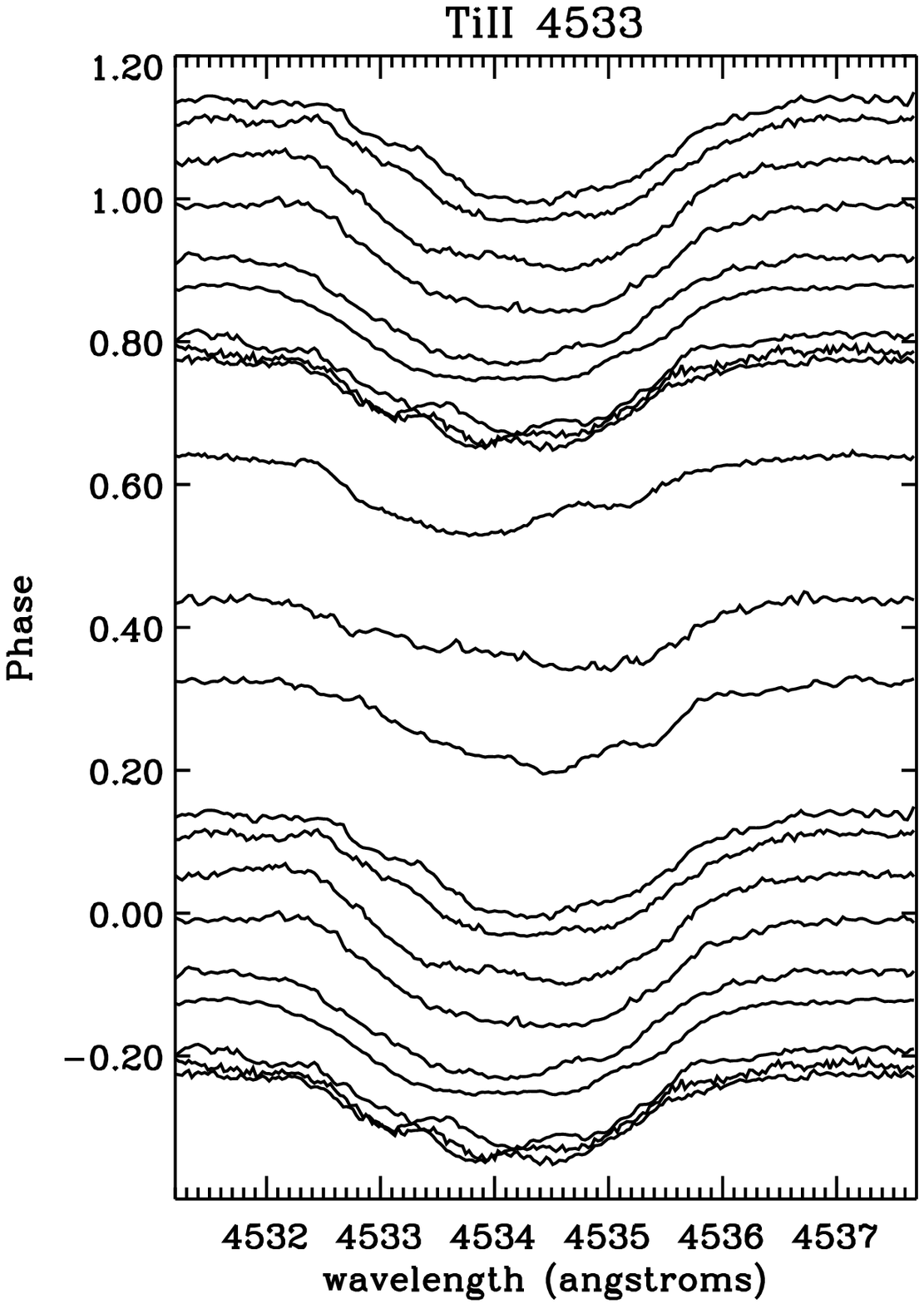}
\includegraphics[width=1.6in]{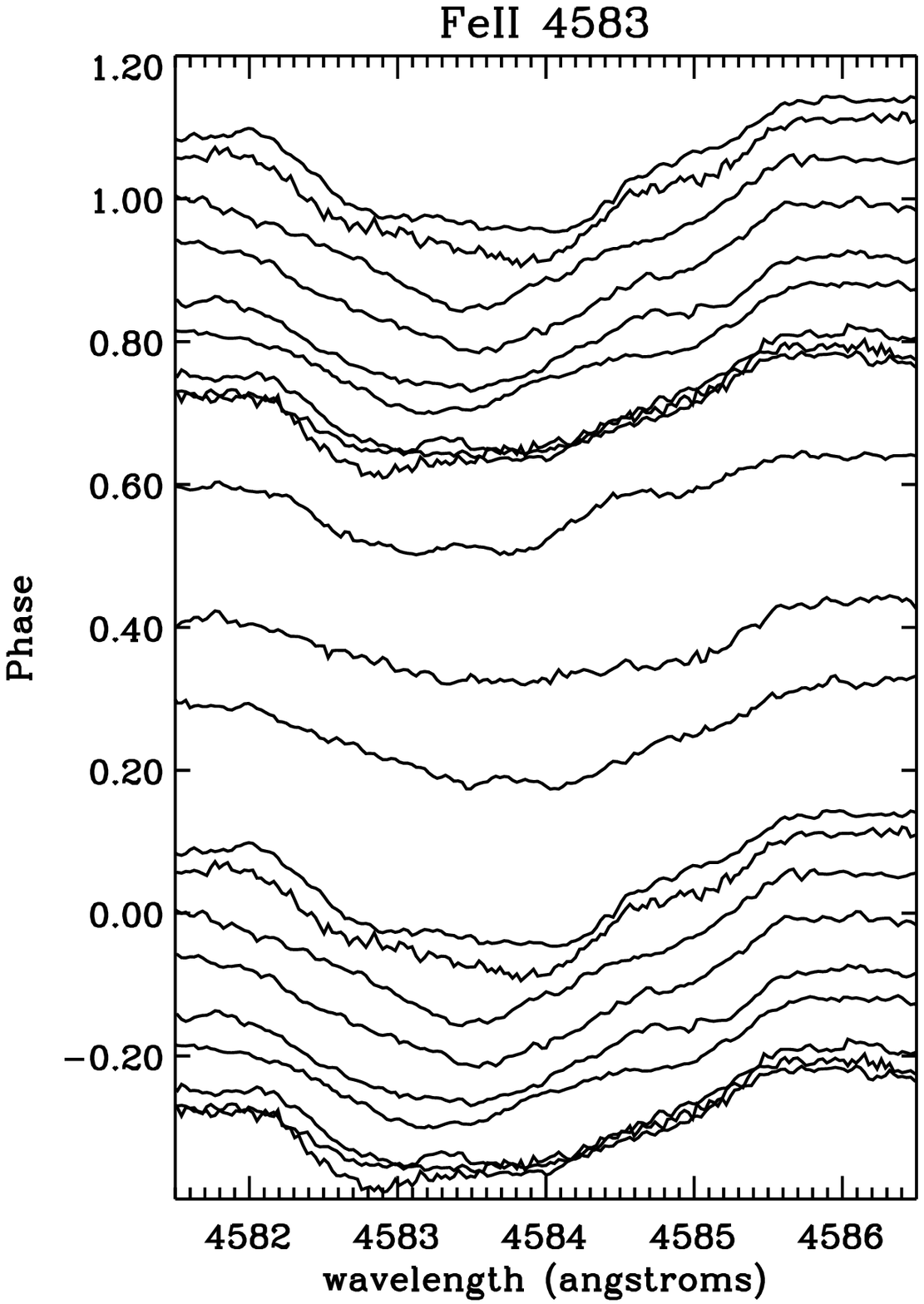}\\
\includegraphics[width=1.6in]{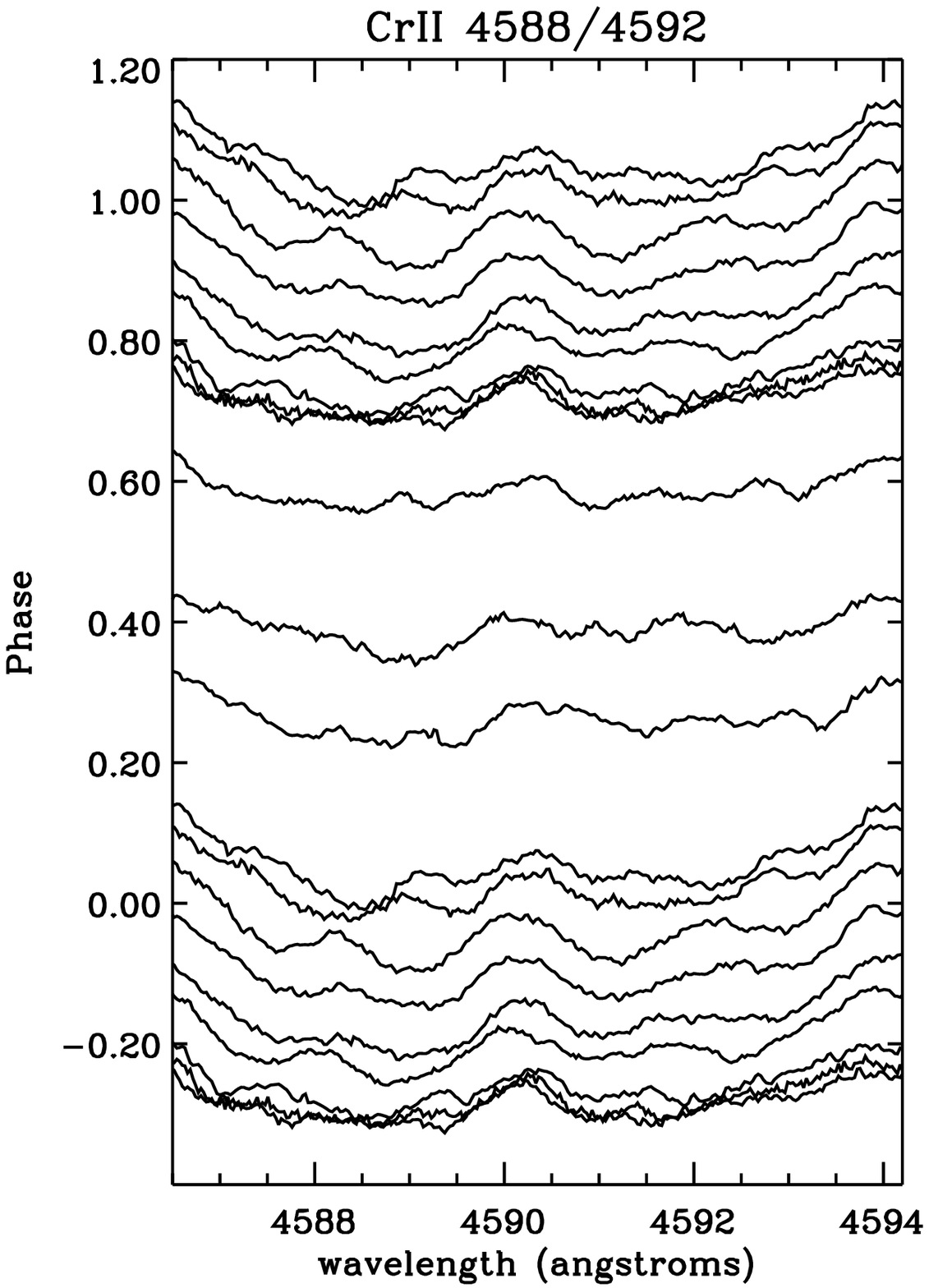}
\includegraphics[width=1.6in]{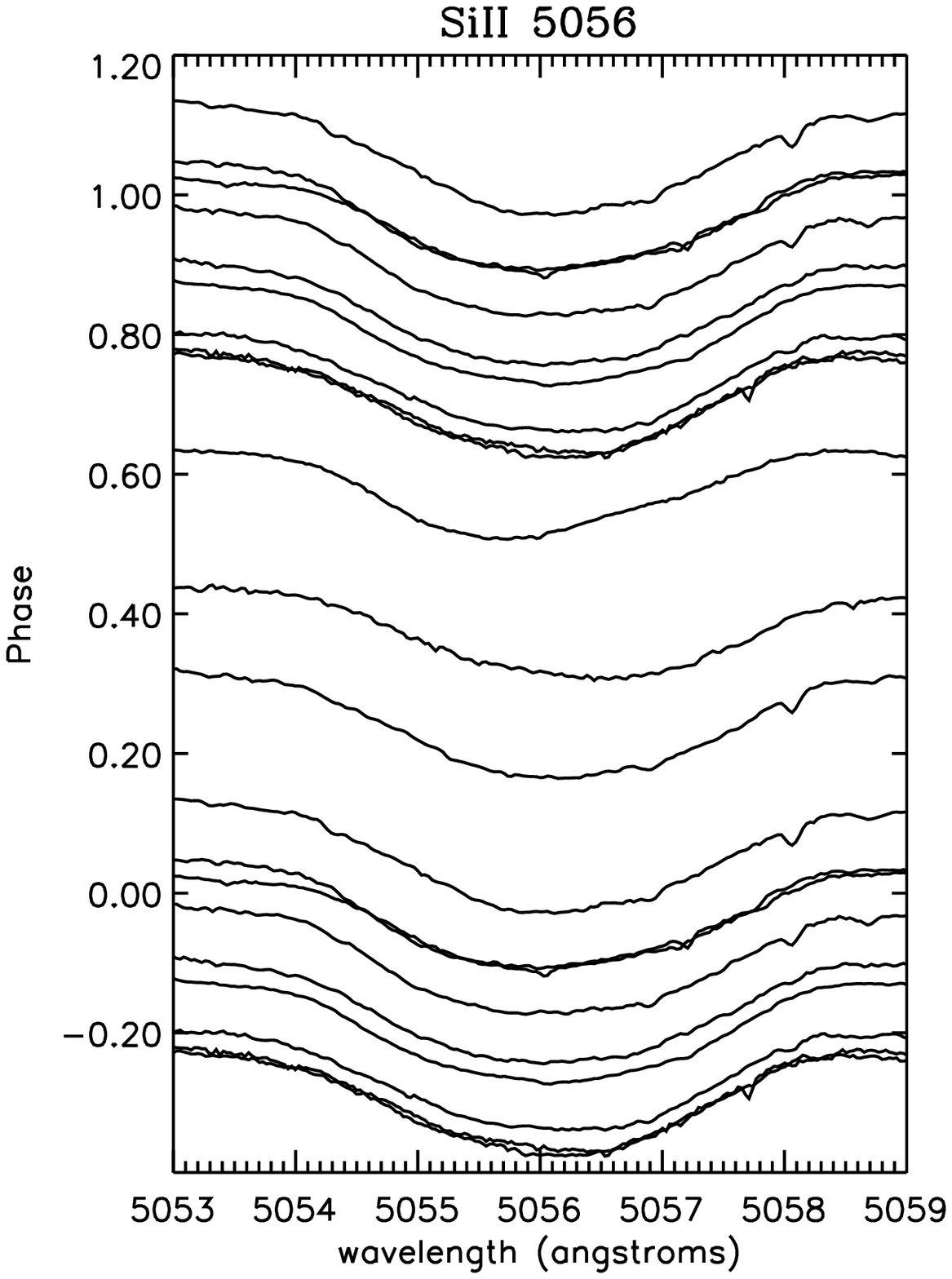}
\caption{Observed metallic line profiles for the indicated lines. Shown are the continuum-normalised spectra, displayed in such a way that the continuum of each spectrum is plotted at a position on the vertical axis that corresponds to the phase of the observation.}
\label{metal_dyn_plot}
\end{figure}

The equivalent width variations for each of the elements plotted in Fig.~\ref{ew_plots} appear to be the result of absorption features travelling through the line profile. The net result is a variation in the total depth, as well as shape, of the line, as demonstrated by the profiles shown in Fig.~\ref{metal_dyn_plot}. The Si line plotted in Fig.~\ref{metal_dyn_plot} has the simplest variations, with what appears to be a single feature travelling from negative to positive velocities. The variations in the the other lines are phased similarly, but the line profiles are generally more complex.

\section{Magnetic Field Model}
\begin{center}
\begin{figure}
\centering
\includegraphics*[angle=-90, width=0.45\textwidth]{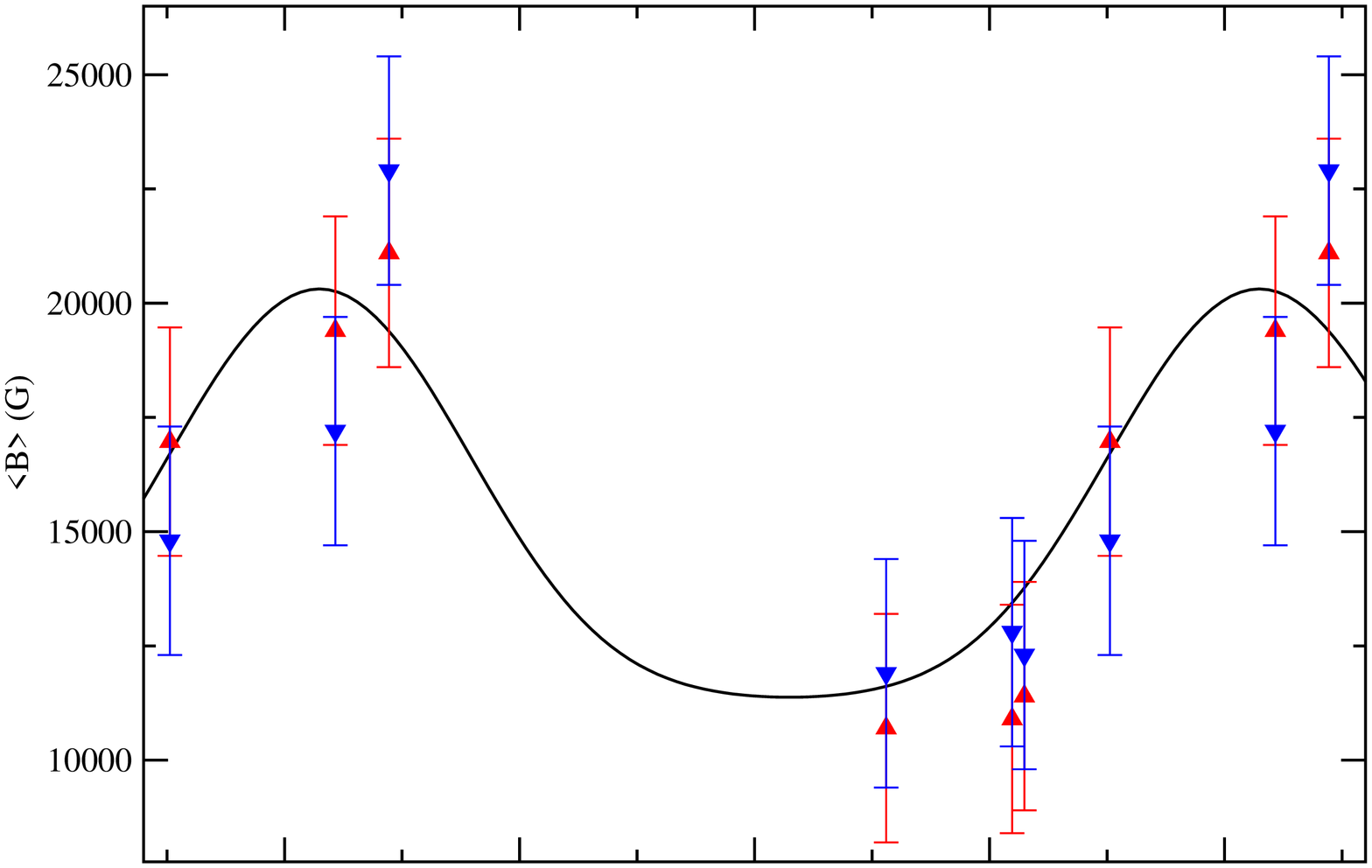}
\includegraphics*[angle=-90, width=0.45\textwidth]{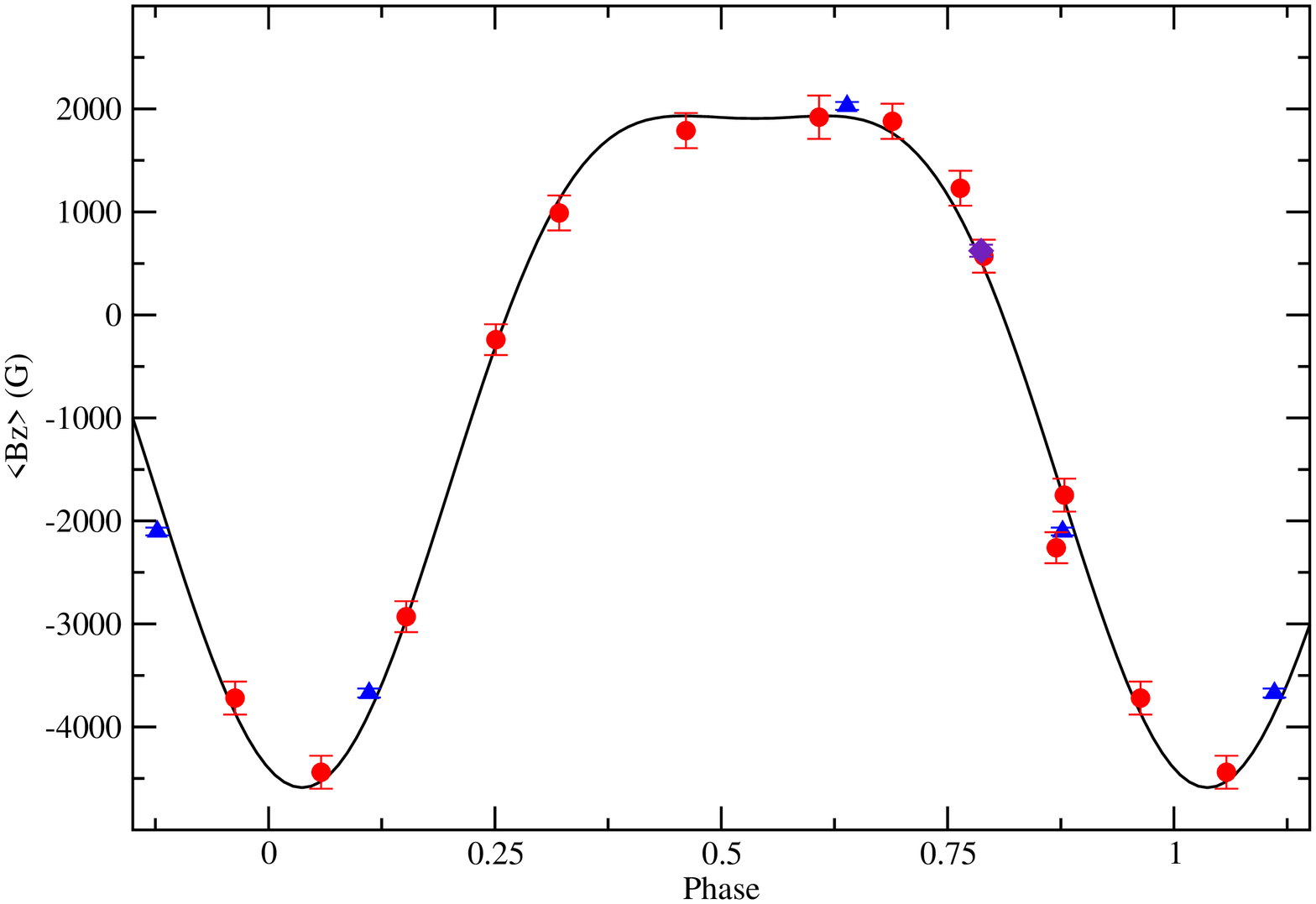}
\caption{The top panel depicts the surface magnetic field modulus variations as measured from Fe~{\sc ii} (red upward facing triangles) and Cr~{\sc ii} (blue downward facing triangles) with the solid (black) curve representing the adopted best-fit model.  The bottom panel shows the \bz\ field variations observed for HD~133880 for our adopted magnetic field model.  The red circles are measurements from \citet{Landstreet1990} using H$\beta$, the blue triangles are LSD measurements from ESPaDOnS, the purple diamond HARPS and the solid (black) curve the adopted magnetic field model.  The minimum of the \bz\ and \bs\ curves occur at phases 0.037 and 0.537 respectively. }
\label{magfield}
\end{figure}
\end{center}
Before any meaningful abundance analysis can be done, we need to establish an appropriate magnetic field model.  \citet{Landstreet1990} derived a best fit magnetic field model based on longitudinal field measurements of H$\beta$ for HD~133880.  He adopted a magnetic field geometry with the inclination angle of the line of sight to the rotation axis, $i$, equal to the angle of the magnetic field axis to the rotation axis, $\beta$ with a value of 90$^{\circ}$ ($i = \beta = 90^{\circ}$).  The dipolar and quadrupolar components were  B$_{d} = -8125$~G, and B$_{q} = -10900$~G respectively.  (Note that we have corrected negative sign errors in both components for the derived field reported by \citet{Landstreet1990}).  From this model, it was established that the predominant magnetic field distribution over the stellar surface is quadrupolar with \bz\ variations that are clearly not sinusoidal.  Further, since a change in the sign of the longitudinal field is observed,  $i + \beta$ must be greater than $90^{\circ}$.  However, Landstreet's adopted geometry is barely consistent with our calculated inclination of $i = 63^{\circ} \pm 18^{\circ}$ (see Sect. 4).  Also, \citet{Landstreet1990} had no information about the variations of  the surface magnetic field modulus, \bs, making it difficult to constrain the field further.  \citet{Landstreet1990} pointed out that there is not a unique model for the observed magnetic curve of HD~133880 and suggested an alternate geometry of $i = 80^{\circ}$ and $\beta = 55^{\circ}$ that produces a similar curve (note that the values of $i$ and $\beta$ can be interchanged without affecting the magnetic curve).  This second geometry is therefore in much better agreement with our derived value of $i$.  Unfortunately, \citet{Landstreet1990} did not provide the values of the dipole and quadrupole components for this alternate geometry.  We therefore search for a new magnetic field model that agrees well with our derived  value for  $i$ of $ 63^{\circ} \pm 18^{\circ}$.   
\subsection{Surface magnetic field modulus measurements}
To further constrain the magnetic field, information about the surface magnetic field modulus \bs\ is required.  To obtain this, we cannot rely on Zeeman splitting because rotational broadening dominates the spectrum, smearing out any observable Zeeman components.  Although dominated by rotation, the magnetic field contributes a measurable fraction to the width of a (Stokes $I$) spectral line.  \citet{Preston1971} outlines a method to obtain mean surface field measurements from rotationally broadened spectral lines.  By comparing the widths of two spectral lines of a given element that have a large and small mean Land\'e factor (ie. lines that are strongly effected by the local magnetic field to ones that are not) we can obtain a meaningful measurement of the magnetic field modulus.  \citet{Preston1971} defines a parameter $K$:
\begin{equation}
K = \left(\frac{w_{L}^{2} - w_{S}^{2}}{\left<\lambda^{4} z^{2}\right>_{L} - \left<\lambda^{4} z^{2}\right>_{S}}\right)^{1/2}, 
\end{equation}   
where $w_{L}$ and $w_{S}$ are the averages of the measured line widths, in cm,  of lines with a large and small Land\'e factor respectively, $\lambda$ is the wavelength in cm and $z$ the mean Land\'e factor where the $L$ and $S$ subscripts refer to the lines with large and small Land\'e factors respectively.  Preston then finds the relationship between the surface field and $K$ to be
\begin{equation}
B_{s}(kG) = 0.5 + 7.9K.
\end{equation}
The challenge with HD~133880 was to find a pair of spectral lines that were not strongly blended with other lines from which to measure line widths.  Table~\ref{bslines} presents two sets of lines for Cr~{\sc ii} and Fe~{\sc ii} from which line widths could be measured.  
\begin{center}
\begin{table}
\caption{Lines used to measure \bs.  The atomic data were gathered from the VALD database.}
\begin{tabular}{ccccc}\hline\hline
 & \multicolumn{2}{c}{Small $z$} & \multicolumn{2}{c}{Large $z$}\\
Element & $\lambda$ (\AA) & $z$ & $\lambda$ (\AA) & $z$ \\ \hline
Fe~{\sc ii} & 4508.288 & 0.50 & 4520.224 & 1.34 \\
Cr~{\sc ii} & 4284.188 & 0.52 & 4261.913 & 1.080\\ \hline
\label{bslines}
\end{tabular}
\end{table}
\end{center}
Although good candidates, these spectral lines are still quite blended with adjacent lines in the spectrum.  To combat this problem, we measured line widths at the continuum by repeatedly fitting Gaussians to the spectral line using IRAF's \textit{splot} function.  The average width for each feature is taken in our calculation.  We were also limited by resolving power and SNR in our FEROS spectra.  Specifically, the combination of the lower resolution of the FEROS instrument and lower SNRs in the data meant that we were unable to unambiguously measure the spectral line widths.  As a result, only the ESPaDOnS and HARPSpol spectra could be utilised.  Results are summarised in Table~\ref{bsmeas} and illustrated in the top panel of Figure~\ref{magfield}.  
\begin{center}
\begin{table}
\caption{\bs\ measurements from lines of Cr~{\sc ii} and Fe~{\sc ii} for spectra from ESPaDOnS and HARPSpol.  Phases are computed from Eq.~2 and uncertainties estimated to be $\pm$2.5~kG.}
\begin{tabular}{cccc}\hline\hline
 &  & \multicolumn{2}{c}{\bs\ (kG)}\\
Instrument & Phase & Fe~{\sc ii} & Cr~{\sc ii} \\ \hline
ESPaDOnS & 0.054 & 19.4 & 17.2 \\
ESPaDOnS & 0.111 & 21.1 & 22.9 \\ 
ESPaDOnS & 0.640 & 10.7 & 11.9 \\
ESPaDOnS & 0.774 & 10.9 & 12.8 \\
HARPSpol & 0.787 & 11.4 & 12.3 \\
ESPaDOnS & 0.878 & 16.9 & 14.8 \\ \hline
\label{bsmeas}
\end{tabular}
\end{table}
\end{center}
It is encouraging that the magnetic field values measured from both Fe~{\sc ii} and Cr~{\sc ii} agree well with one another, suggesting the variations are indeed real.   Due to the high blending in the spectral lines and the ambiguity in measuring the line widths, we estimate the uncertainties to be $\pm$ $\sim$  2500 G, though this value may be optimistic.  Although this method is not as precise as measuring the splitting of spectral lines due to the Zeeman effect, it does provide meaningful constraints on the magnetic field.  It is clear from these measurements that the surface field varies from about 20~kG at the negative magnetic pole to 10~kG near the positive magnetic pole (see below).  The derived field is consistent with a non-sinusoidal curve (though no definitive conclusions can be made given the limited number of data points and size of the error bars), much in the same manner to the line-of-sight magnetic field, but in the opposite sense:  the maximum of the longitudinal field curve corresponds to the minimum of the surface integrated field curve.  More measurements of \bs\ are required to constrain the field further which requires more high SNR observations of HD~133880.    

\subsection{Magnetic field geometry}
\begin{center}
\begin{figure*}
\centering
\includegraphics[width=0.45\textwidth]{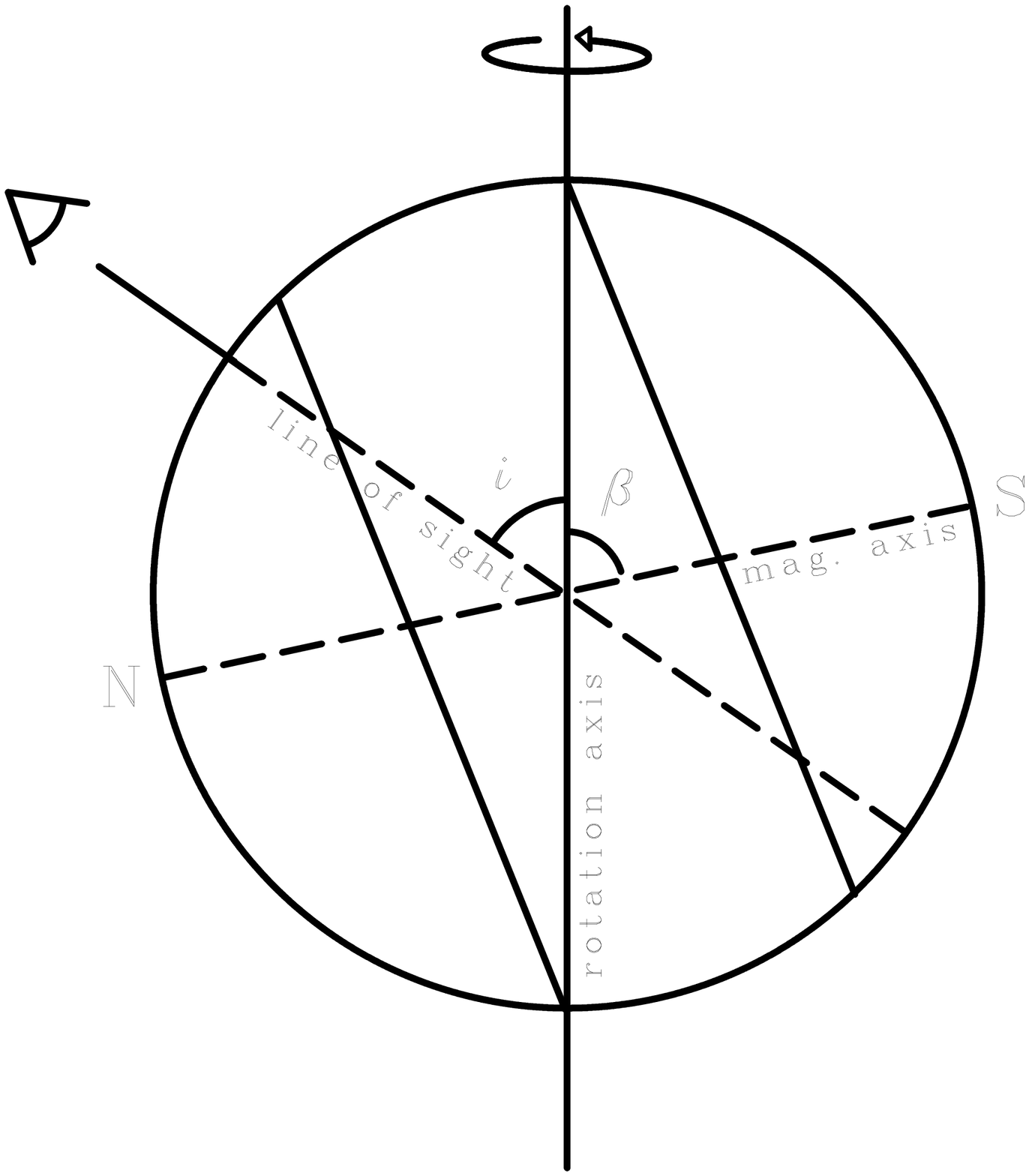}
\includegraphics[width=0.45\textwidth]{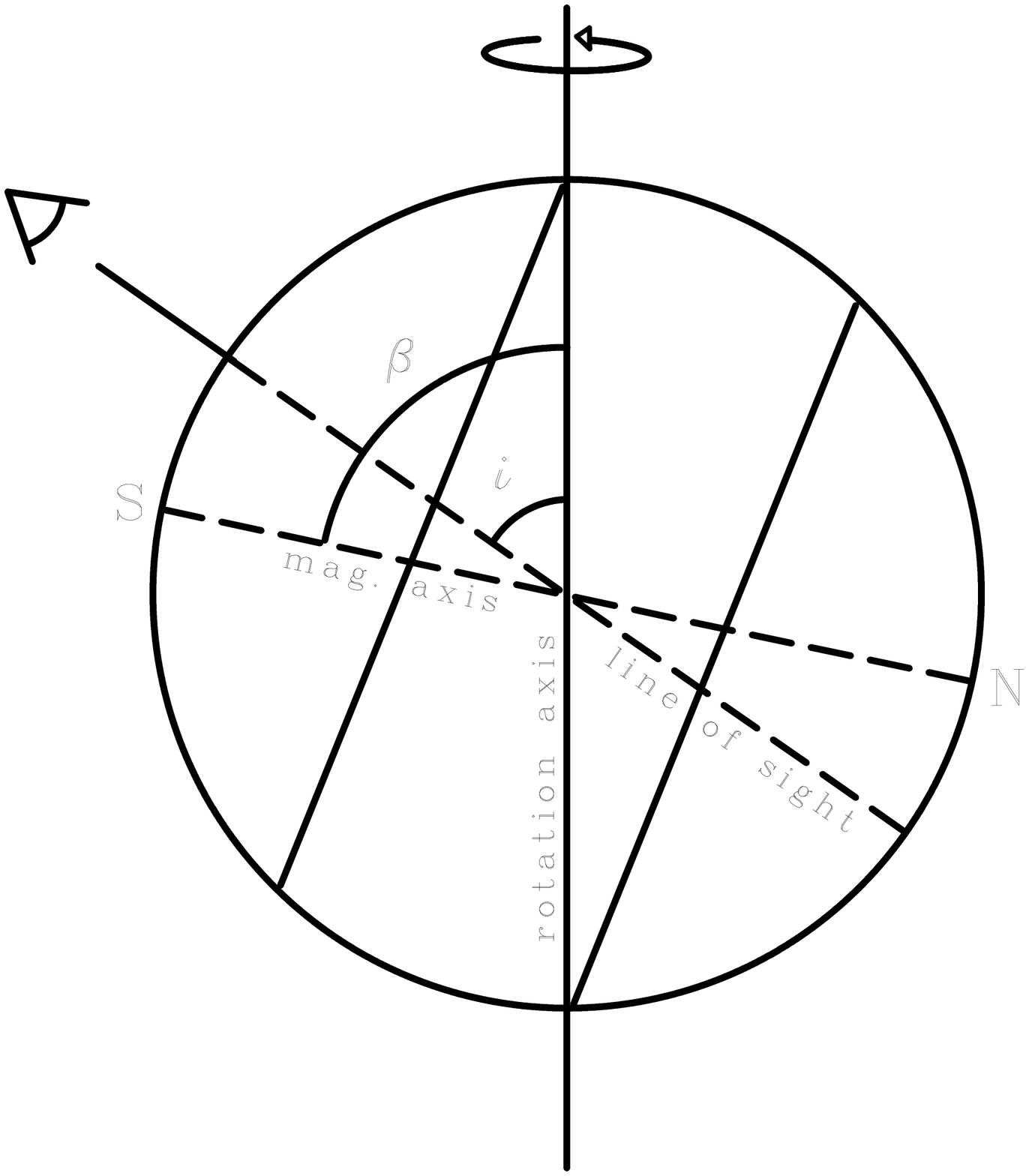}
\caption{The geometry adopted for HD~133880.  The vertical axis is the rotation axis of the star.  The angle between the line of sight and the rotation axis is $i = 55^{\circ}$.  The angle between the rotation axis and the magnetic field axis is $\beta = 78^{\circ}$.  The left panel depicts an orientation where the positive magnetic pole is closest to the line of sight ($\phi = 0.537$) and the right panel is when the negative magnetic pole is nearer alignment with the line of sight ($\phi = 0.037$).  The bands used for dividing the abundance distribution are shown as solid lines and are normal to the magnetic field axis.}
\label{oblique}
\end{figure*}
\end{center}   
The FLDSRCH.F programme was used to obtain the magnetic field geometry for HD~133880 \citep{LM2000}.  This programme has longitudinal and surface magnetic field strengths at four different phases as input, which are then used to iteratively search for the best fit $i$, $\beta$, B$_{d}$, B$_{q}$, and B$_{oct}$.  For a more complete description refer to \citet{Bailey2011}.  Measurements  of \bz\ from \citet{Landstreet1990}, as well as the four new measurements from this work, provide good constraints on the observed variations.  We also have important limits on \bs\ that appear to range from 10~kG to 20~kG near the positive and negative magnetic poles respectively (see above).  This information allowed us to derive a new field geometry: $i$=55$^{\circ} \pm 10^{\circ}$, $\beta$=78$^{\circ} \pm 10^{\circ}$, $B_{d} = -9600 \pm 1000$~G, $B_{q} = -23200 \pm 1000$~G, and $B_{oct} = 1900 \pm 1000$~G.  Uncertainties were estimated by observing the deviations from the derived field by changing the input parameters of \bz\ and \bs\ within their uncertainties.  This field corresponds well with our calculated value of $i = 63^{\circ} \pm 18^{\circ}$ and has the added advantage of agreeing with the alternate geometry proposed by \citet{Landstreet1990} (see above).  An illustration of our adopted geometry is presented in Figure~\ref{oblique}.  The bottom panel of Figure~\ref{magfield} shows the model \bz\ curve over-plotted with \bz\ measurements from \citet{Landstreet1990} and this paper.  All the measurements agree well with the computed model and the measurements made using the metallic spectrum agree remarkably well with \citet{Landstreet1990} who made measurements from H$\beta$.  The field varies non-sinusoidally from about -4.5~kG at phase 0.037 to 2~kG at phase 0.537.  The field curve is flat and extended around the positive magnetic pole with a rapid decline towards the negative magnetic pole.  The quadrupole-to-dipole ratio, $q$, of the adopted model is 2.4, which clearly indicates the presence of a strong quadrupole component.  We therefore concur with the conclusion of \citet{Landstreet1990} that the field of HD~133880 departs in a significant way from a simple centred dipole.

\section{Abundance Model}
We utilise the FORTRAN programme ZEEMAN \citep{L1988, L1989, wade2001} to perform abundance analysis of HD~133880.  ZEEMAN computes emergent spectra of all four Stokes parameters for stars with a permeating magnetic field.  It assumes a low-order multipole geometry for the magnetic field based on the above derived values for $i$, $\beta$, B$_{d}$, B$_{q}$, and B$_{oct}$.  Up to ten phases of observed Stokes $I$ spectra can be fit simultaneously with the abundance of one element varied at a time.  For each iteration, ZEEMAN automatically adjusts \vs\ and the radial velocity of the star to optimum values.  Currently, up to six rings of uniform abundance on the surface of the star can be specified that have equal spans in magnetic co-latitude.  ZEEMAN produces output for the abundance distribution of a given element at different magnetic latitudes.  In effect, this model provides a rough 1D map of the variation of abundance from one magnetic pole to the other.

Initially, uniform mean abundances for each phase of HD~133880 were found, however, we quickly discovered that a multi-ring model would provide a better fit.  As explained by \citet{Bailey2011}, ZEEMAN was designed to quantify hemispherical abundance variations for stars similar to 53~Cam (=HD~65339) \citep{L1988}.  HD~133880 exhibits large-scale abundance variations over the stellar surface and, as is evident from the longitudinal magnetic field curve, both magnetic hemispheres are observed.  Our preliminary abundance analysis suggested that a maximum of three rings should be used to quantify the abundance variations because we saw abundance anomalies that varied between both magnetic poles and the magnetic equator.  Nevertheless,  we experimented with fewer and more rings on the surface of the star.  We found that more than three rings did not significantly improve the quality of fits and fewer rings did not adequately model the observed variations in the Stokes $I$ spectra.

For our analysis, \teff\ was set to 13000~K and \lgg\ to 4.3 which are both consistent with the values determined in Sect. 4.  The atomic data were taken from Vienna Atomic Line Database \citep{vald3, vald2, vald4, vald1}.  The best-fitting value of \vs\ was found to be $103 \pm 10$~\kms.  Due to the strong blending in the spectral lines, identifying the best fit required visual evaluation of the agreement between the computed and the observed Stokes $I$ spectral lines in several windows of the order of 80~\AA\ wide.  Since the strong magnetic field should suppress any convective motions, which in any case are expected to be small at this \teff, the microturbulence parameter was set to 0~\kms.  

\subsection{Spectra normalisation}
For stars in which \vs\ is greater than around 100~\kms\ (as is the case for HD~133880) the dominant source of error in abundance analysis is from continuum normalisation.  The reason for this is that in a spectral region that is on the order of 100~\AA\ wide, there may only exist a few continuum points from which to normalise the spectrum.  Often it is difficult to determine which features should be placed near the continuum and which should not.  To combat this problem, we performed continuum normalisation of spectra of normal, sharp-lined, A-type stars from both the ESPaDOnS and FEROS instruments to determine the highest degree polynomial required to normalise a given spectral window.  All spectral windows were about 100~\AA\ wide to maximize the number of continuum points present in the broad-line spectra.  In all cases, no greater than two segments of a cubic spline were required to normalise the normal A-star spectra correctly.  For a given instrument and spectral window, we also used two segments of a cubic spline to continuum fit the spectra for HD~133880.  In this manner, we reduce the uncertainty introduced in the placement of the continuum and ensure that all spectra are normalised in the same way.  Where necessary, we iteratively performed continuum fitting for spectral windows for which synthetic spectra from ZEEMAN agreed poorly with our initial normalisation.  \citet{hill1995} discusses further challenges in performing abundance analysis on broad-lined stars. 

\subsection{Choice of magnetic field model}
As described in Section 6, two magnetic field models adequately characterise the observed variations of the longitudinal magnetic field, \bz: that of \citet{Landstreet1990} and the one presented in this paper.  To explore the effect of different magnetic field geometries on the derived abundances, we performed abundance analysis of HD~133880 for all elements using both geometries.  The maximum abundance differences between the two models were found to be 0.1-0.2~dex.  We therefore adopt the new magnetic field geometry presented in this paper because it agrees well with the value of $i$ derived from physical parameters and has the added constraints of the observed surface magnetic field modulus variation, \bs.  We report all abundances using the new magnetic field geometry. 

\section{Abundance analysis}
\begin{center}
\begin{table*}
\caption{Abundance distribution of elements studied.  Only upper limits of He, Ni and Nd are found.}
\centering\begin{tabular}{ccccccccccc}
\hline\hline & \multicolumn{10}{c}{Log($n_{\rm X}$/$n_{\rm H}$)}\\
  & He & O & Mg & Si & Ti & Cr & Fe & Ni & Pr & Nd\\\hline
Ring $0-60^{\circ}$ (negative pole) & $\leq -1.90$  & -2.78 & -3.84 & -2.79 & -5.39 &-4.55 & -3.32 & $\leq -4.40$ &-6.52 & -6.55 \\
Ring $60-120^{\circ}$ (magnetic equator)  & -- & -3.38 & -4.37 & -2.64 & -5.62 &-4.54 & -3.44 & -- &-6.72 & -6.47 \\
Ring $120-180^{\circ}$ (positive pole) & -- & -3.10 & -3.91 & -3.93 & -6.75 & -5.20 & -4.11 & -- & -7.28 & -8.22 \\
$\sigma$ & -- & $\pm$0.3 & $\pm$0.3 & $\pm$0.2 & $\pm$0.2 & $\pm$0.15 & $\pm$0.1 & -- & $\pm$0.2 & $\pm$0.3 \\
\\
Solar abundance & -1.07 & -3.31 & -4.40 & -4.49 & -7.05 & -6.36 & -4.50 & -5.78 & -11.28 & 
-10.58 \\
\\
\# of lines modelled & 2 & 1 & 1 & 4 & 3 & 3 & 3 & 1 & 3 & 1 \\
\hline\hline
\label{abunddistr}
\end{tabular}
\end{table*}
\end{center}  
A total of 12 spectra, well spread in phase, were available for HD~133880, from the FEROS, ESPaDOnS and HARPS instruments.  The HARPS spectrum was not used in the abundance analysis because its phase corresponded closely to two other spectra with comparable SNR and thus added no new relevant information.  The large wavelength coverage of these instruments has allowed us to derive abundance distributions for O, Mg, Si, Ti, Cr, Fe, and Pr, as well as upper limits for the abundances of He, Ni, and Nd.  The large \vs\ presented challenges in deriving abundances due to the presence of significant blending in the spectra.  Nevertheless, multiple relatively clean lines of the majority of elements studied were found with varying strengths and Land\'e factors, providing better constraints on the derived abundances.   

As discussed in the previous section, the present dataset is sufficient to perform abundance analysis using a three-ring model with rings encompassing both magnetic poles and the magnetic equator with equal span in co-latitude ($60^{\circ}$).  Initially six spectra were used (2 ESPaDOnS and 4 FEROS) that were well spaced in phase to derive the three-ring abundance model.  Once the mean abundance values were found for each ring, model fits were produced for all spectra.  We have no observations for HD~133880 very close to the positive magnetic pole (phase 0.537).  We do, however, have two spectra sufficiently near the positive pole, at phases 0.438 and 0.640, adequate for providing a first approximation to the abundance distributions near the positive pole.

The mean abundances derived for each element in each ring are tabulated in Table~\ref{abunddistr} along with solar abundances reported by \citet{solar}.  The quality of fits were tested (where possible) using multiple spectral windows with derived uncertainties estimated from the observed window to window scatter of the abundance.  For lines for which there was only one spectral line or region, we estimate the uncertainties to be about $\pm$0.3 dex, which is the average deviation from the derived abundance required to make the quality of fit unsatisfactory between the model and observed spectra upon visual examination.   We also derived a set of abundances using \teff\ = 12000~K as reported by \citet{netopil2008} for each of the elements listed in Table~\ref{abunddistr}.  We found that the abundances in each of the three rings changed by $\pm$0.10~dex, or less, for the Fe-peak elements (Ti, Cr, Fe) and no more than $\pm$0.15~dex for the rare-earths (Pr and Nd), O, Mg, and Si which are within our quoted uncertainties.  This suggests that the choice of \teff\ does not significantly influence the derived abundances for HD~133880.  To illustrate the quality of fits of the three-ring model, two 80~\AA\ windows for all 11 modelled spectra are shown in Figures~\ref{500-508} and \ref{453-460}.  The quality and consistency of the model are discussed below for each element modelled. 

\subsection{Helium}
Many possible lines of He~{\sc i} are considered when determining the abundance of helium:  4471, 5015, 5047, 5876, and 6678~\AA.  However, none of these lines are unambiguously detected.  Nevertheless, fitting these regions provides a useful upper limit to the abundance of helium.  Our analysis confirms that HD~133880 is a He-weak star with the abundance of helium at least a factor of 8 below the solar abundance.

\subsection{Oxygen}
HD~133880 is classified as a Bp star with the $\lambda$4200-Si anomaly which suggests that the oxygen abundance should be near or less than the solar abundance \citep{oxygen1962,oxygen1990}.  Initially, we attempted to derive the oxygen abundance using the 7771-75~\AA\ triplet.  These lines suffer from strong non-LTE effects and are highly saturated, making them unreliable for abundance determination.  A more useful set of lines are those at 6155-56-58~\AA; the abundance distribution reported in Table~\ref{abunddistr} are from these lines.  These lines are unambiguously detected, but are somewhat blended with Pr~{\sc ii} near the redward wing. The O abundance varies from the magnetic poles to equator by a factor of about 2-3.  At all phases, oxygen is found from these lines to be overabundant near the negative and positive magnetic poles by 0.5~dex and 0.3~dex compared to the solar ratio, respectively.  Near the magnetic equator, oxygen is comparable in abundance to the Sun, within uncertainties.

\subsection{Magnesium}
Only one clean line of sufficient strength to model was found for Mg~{\sc ii} at 4481~\AA.  This line shows some variability, implying an overabundance of around 0.5~dex near both magnetic poles and a roughly solar abundance near the magnetic equator.

\begin{center}
\begin{figure*}
\centering
\includegraphics[angle=-90,width=0.95\textwidth]{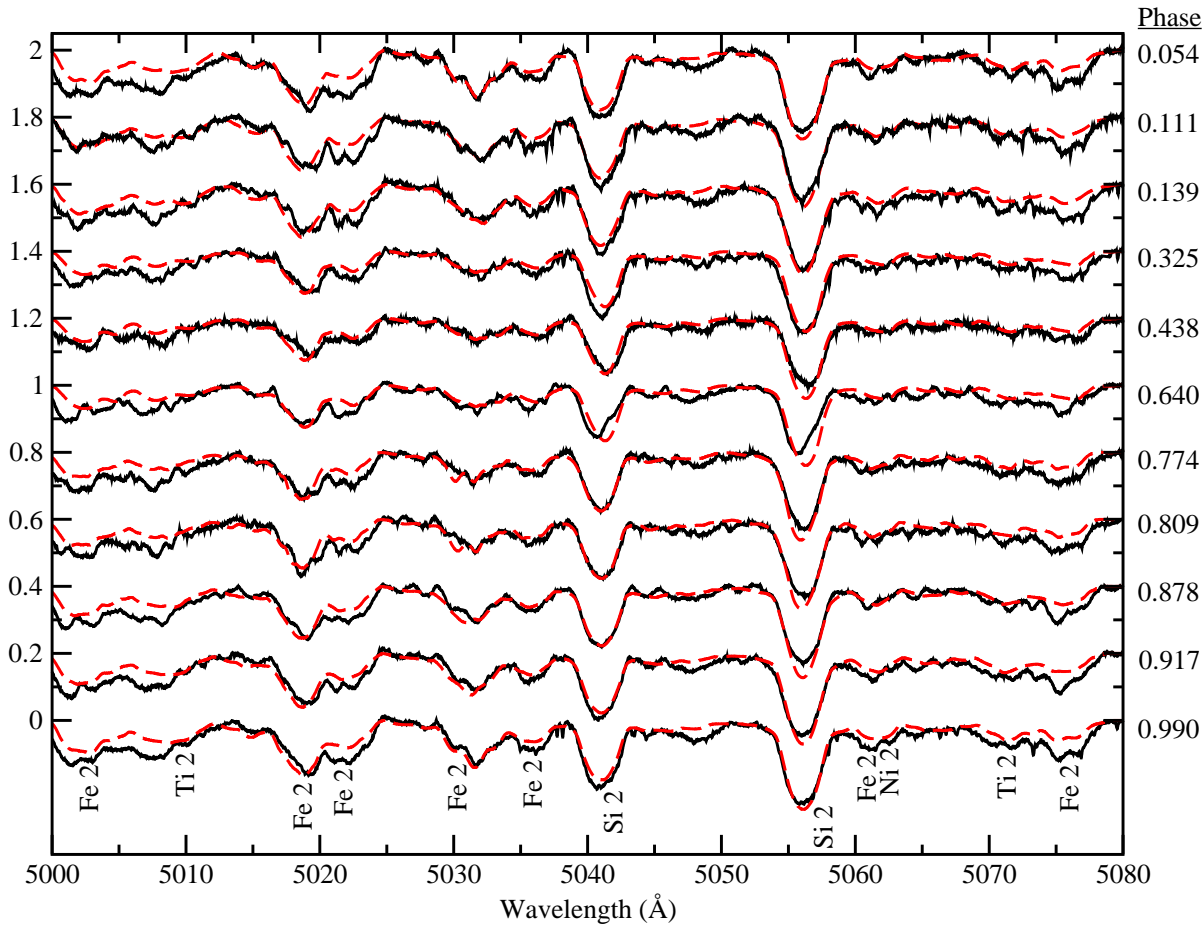}
\caption{Spectrum synthesis of the region 5000 - 5080~\AA\ using a three-ring model.  The observed spectra are in black (solid) and the model fits are in red (dashed).  From top to bottom phases 0.054, 0.111 (both ESPaDOnS), 0.139, 0.325, 0.438 (all FEROS), 0.640, 0.774 (both ESPaDOnS), 0.809 (FEROS), 0.878 (ESPaDOnS), 0.917, and 0.990 (both FEROS).  The line of sight is closest to the negative magnetic pole at phase 0.037 and the positive magnetic pole at phase 0.537.  Spectra are positioned at arbitrary positions along the vertical axis for display purposes.}
\label{500-508}
\end{figure*}
\end{center}

\subsection{Silicon}
A total of four lines of Si~{\sc ii} were found that were suitable for modelling: 5041~\AA, the doublet at 5055-56~\AA, 5955~\AA\, and 5978~\AA.  The first two lines of Si~{\sc ii} were fit simultaneously and the result tested by using the same derived abundance model in the latter two lines.  Satisfactory fits were obtained at each phase for all lines of Si, however the doublet at 5055-56~\AA\ was systematically too strong in the model at all phases (see Figure~\ref{500-508}).  Finding the best fit model for the longer wavelength lines did not rectify the problem of stronger spectral lines in the model and produced a similar abundance distribution.  We find large-scale abundance variations for Si on the surface of HD~133880 from pole to pole.  At the negative magnetic pole and magnetic equator, Si is approximately 25 times more abundant than in the Sun.  The lowest abundance is found at the positive magnetic pole where, however, Si is still about 3 times more abundant than in the Sun. 

  Figure~\ref{453-460} highlights three Si~{\sc iii} lines at 4552, 4567, and 4574~\AA\ that are not modelled well using the abundance derived from the Si~{\sc ii} lines discussed above (red-dashed line).  An enhancement of the Si~{\sc ii} abundance of about 1~dex is necessary to model well the Si~{\sc iii} lines (blue-dashed line).  This discordance is very suggestive.  The effects of non-LTE in the Si~{\sc iii} lines are expected to be small at \teff\ less than about 15000~K.  \citet{beckerandbutler1990} demonstrate that non-LTE effects in the Si~{\sc iii} lines decrease towards lower temperatures.  The most probable cause is strong stratification of Si in the atmosphere of HD~133880 with the abundance high in the atmosphere being much lower than near an optical depth ($\tau_{c}$) of about 1.  However, the large \vs\ (103~\kms) makes normalisation difficult (see Sect. 7.1) and studies on more hot Ap stars with low \vs's are necessary to determine how widespread this Si~{\sc ii}/{\sc iii} discrepancy is, and how it may be explained.

\subsection{Titanium}
We have modelled the abundance of Ti using lines of Ti~{\sc ii} at 4563 and 4571~\AA\ as well as at 4805~\AA.  The final model was determined using the former two lines and tested using the latter line.  Ti exhibits drastic variations between magnetic poles of nearly 1.5~dex which can be seen in Figure~\ref{453-460} where lines due to Ti~{\sc ii} are much weaker at phases 0.423 and 0.625 than near phase 0.  Ti is most abundant at the negative magnetic pole where it is more than 40 times more abundant than in the Sun and transitions to slightly lower abundances at the magnetic equator ($\sim$ 25 times overabundant).  At the positive magnetic pole Ti is only about 2 times more abundant than the solar value. 

\subsection{Chromium}
\begin{center}
\begin{figure*}
\centering
\includegraphics[angle=-90,width=0.95\textwidth]{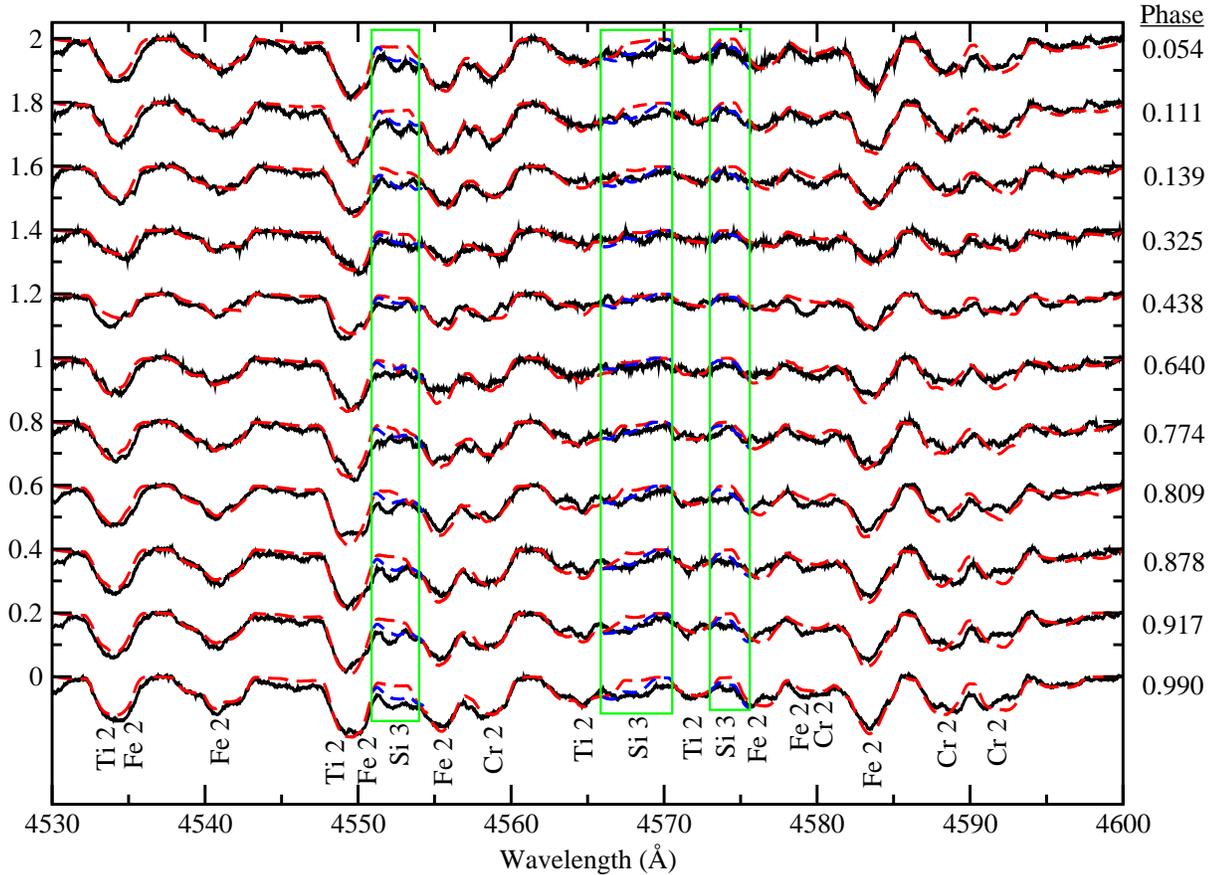}
\caption{Same as Figure~6 for the region 4530 - 4600~\AA.  The green boxes highlight the three Si~{\sc iii} lines at 4552, 4567, and 4574~\AA.  The red-dashed lines indicate the model fits to abundances derived from the Si~{\sc ii} lines that are presented in Table~\ref{abunddistr}.  The blue-dashed lines are model fits with about a 1~dex enhancement to the Si abundances shown in Table~\ref{abunddistr}.}
\label{453-460}
\end{figure*}
\end{center}

Several lines of Cr were found that were suitable for modelling, particularly the Cr~{\sc ii} lines at 4558, 4588, and 4592~\AA.  The latter two of these three lines were relatively clean while $\lambda$4558 is blended with Fe~{\sc ii}.  All three lines were used to compute the three-ring model presented in Table~\ref{abunddistr}.  All Cr lines in Figure~\ref{453-460} are fit well at all phases with the model fit at 4592~\AA\ being slightly too strong at later phases.  Large variations are observed from hemisphere to hemisphere.  Cr is weakest at the positive pole, but its abundance is still 10 times greater than the solar abundance.  The negative pole and magnetic equator have a mean abundance that is approximately 100 times greater than in the Sun.  

\subsection{Iron}
A plethora of lines are available from which to derive the abundance distribution of Fe.  The model was found using Fe~{\sc ii} lines at 4541, 4555, and 4583~\AA.  The lines were of sufficient strength for abundance analysis and only the line at 4555~\AA\ is blended with Cr~{\sc ii}.  Similar to Cr, Fe is least abundant at the positive magnetic pole, where an abundance of about 2.5 times higher than the solar abundance is found,  and most abundant near the negative magnetic pole and magnetic equator where a mean abundance approximately 10 times larger than in the Sun is observed.  Most lines of Fe are modelled well, as seen in Figure~\ref{453-460}.  Some weaker lines are systematically too weak in the model (see Figure~\ref{500-508}) which is most likely attributable to unrecognized blending.

\subsection{Nickel}
There are a few possibly useful lines of Ni~{\sc ii} to model at 5058, 5059, and 5064~\AA, however none are unambiguously detected.   Nevertheless, the region (labelled in Figure~\ref{500-508}) proved useful for obtaining an upper limit to the abundance of Ni, which suggests that it is less than 10 times more abundant than in the Sun.

\subsection{Praseodymium}
Three lines of Pr were found that were useful for modelling.  The final model was computed using Pr~{\sc iii} and Pr~{\sc ii} at 6160 and 6161~\AA\ respectively.  These lines are blended with the redward wing of the O~{\sc i} triplet at 6155-56-58~\AA, but are of sufficient strength to obtain a useful abundance.  The model fits were then tested using Pr~{\sc iii} at 7781~\AA.  A mean abundance was found for the two rings covering the negative magnetic pole and magnetic equator that is about 10$^{5}$ times overabundant compared to the solar ratio.  At the positive magnetic pole, Pr is least abundant but still 10$^{4}$ times more abundant than in the Sun.
    
\subsection{Neodymium}
A possible useful line of Nd~{\sc iii} is at 6145~\AA, however the SNR is such that it is not unambiguously detected at all phases and rotational broadening causes this line to be strongly blended with Fe~{\sc ii} at 6147 and 6149~\AA.  We therefore turn to Nd~{\sc iii} at 4711-12-14~\AA.  These lines are relatively clean, providing useful abundance determinations that reveal a strong variation in the mean abundance of Nd between magnetic hemispheres.  For both the negative magnetic pole and magnetic equator Nd is greater than 10$^{4}$ times more abundant than in the Sun, whereas at the positive pole this rare-earth is about 200 times the solar ratio.  

\section{Magnetosphere}

\citet{Lim1996} demonstrated that the 3.5 cm and 6 cm radio flux and circular polarisation of HD 133880 vary significantly and coherently according to the $\sim 0.877$~d period. They reported that the emission shows broad peaks near the phases of the longitudinal field extrema (corresponding, in Landstreet's model, to the phases at which the poles of the dipole contribution to the field curve pass closest to the line-of-sight), and narrower peaks at the predicted phases of quadrupole component pole passages (i.e. the quadrupole contribution to the field curve). They also reported that the strong extraordinary-mode circular polarisation varied in sign with phase, in agreement with the sign of the longitudinal field.  As described in Sect.~2, we have re-reduced the ATCA radio data, but there are no significant differences in the reduced data compared to that described by \citet{Lim1996}. Here we examine the radio emission in the context of a magnetospheric picture combining the dynamical concepts summarised by \citet{owocki2006} and the radio magnetosphere model of \citet{linsky1992}.

Using the stellar wind parameters determined using the CAK formalism (\citet{castor1975}: values of $\dot M=10^{-11}~M_\odot$/yr and $v_\infty=750$~\kms), we compute the wind magnetic confinement parameter $\eta_*=B_{\rm eq}^2\,R^2/\dot M v_{\rm \infty}\simeq 10^7$  (neglecting clumping, \citet{udDoula2002}), where $B_{\rm eq}$ is the equatorial surface strength of the magnetic dipole component.  The rotation parameter $W=v_{\rm eq}/v_{\rm crit}=0.3$ (\citet{udDoula2008}, where we have included a correction for the oblateness of the star due to its rapid rotation). This places the Alfv\'en radius at $R_{\rm Alf}=\eta_*^{1/4}~R_* \simeq 60~R_*$. In contrast, the Kepler (or corotation) radius is located relatively close to the star, at about $R_{\rm Kep}=W^{-2/3}~R_*=2.2~R_*$.  This geometry, in which $R_{\rm Kep}<<R_{\rm Alf}$, results in a large spatial volume in which magnetospheric plasma is predicted to be maintained in rigid rotation with the star by the magnetic field, while simultaneously being dynamically supported against gravitational infall due to rapid rotation \citep{udDoula2008}. 

In principle, this situation provides conditions suitable for significant accumulation of rigidly-rotating stellar wind plasma (diagnosed from optical and UV line emission and variability; e.g. \citet{Petit2011}), as well as a large region over which electrons can be accelerated (thus producing the non-thermal radio emission; e.g. \citet{linsky1992}).

\begin{figure}
\centering
\includegraphics[angle=-90,width=3.3in]{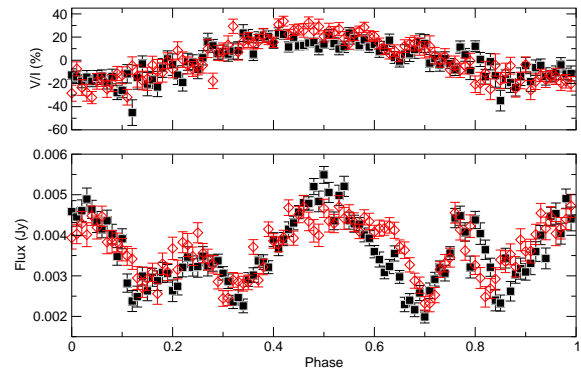}\\
\caption{{\bf Lower frame:} ATCA Stokes $I$ flux variation at 6 cm (filled symbols) and 3 cm (open symbols). {\bf Upper frame:} Stokes $V/I$ polarisation at 6 cm (filled symbols) and 3 cm (open symbols).}
\label{radio_plot}
\end{figure}

The phased radio light curves, illustrated in Fig.~{\ref{radio_plot}}, exhibit a number of interesting characteristics. First, it is clear that the radio measurements in Stokes $I$ and $V/I$ vary significantly, and are coherently phased with the rotational period derived from the photometric and magnetic measurements. Comparison of the 3.5 cm and 6 cm polarisation and light curves fully confirm the details of the variations. As described by Lim et al., the flux variation is complex, characterised by strong, broad maxima at phases 0.0 and 0.5 (i.e. the extrema of the longitudinal field), and sharper, somewhat weaker secondary extrema at quadrature phases (i.e. 0.25 and 0.75). The polarisation degree varies approximately sinusoidally, with extrema of $\pm 16$\% and a mean of zero. The new rotational period derived in Sect.  3 brings the phases of the radio flux and polarisation extrema into good agreement with the longitudinal field and photometric extrema, resolving the $\sim 0.05$-cycle offset pointed out by \citet{Lim1996}. The approximate reflectional symmetry of the light curve in both $I$  and $V/I$ (horizontally, relative to phase 0.0, as well as vertically for Stokes $V/I$, relative to the $V/I=0$) is in good qualitative agreement with the derived stellar geometry, in particular the large value of the magnetic obliquity $\beta\simeq 90^{\rm o}$, implying that we view geometrically similar regions of both (magnetic) hemispheres during a stellar rotation (see also Fig. 5).

Comparing the variations at 3.5 and 6 cm, we find that the amplitudes are nearly identical in both $I$ and $V/I$. The Stokes $I$ variations may display some small differences in shape, in particular with the central (phase 0.5) peak at 3.5 cm being somewhat shallower and broader than at 6 cm.

The presence of structures in the radio lightcurve attributable to the quadrupolar component of the field is very interesting. While the Alfv\'en radius for the dipole component fall off as $\eta_*^{1/4}$, for the quadrupole the fall-off is as $\eta_*^{1/6}$ (Ud Doula, Owocki and Townsend 2008). The ``quadrupolar Alfv\'en radius'' for HD 133880, computed assuming the magnetic model derived in Sect. 6.2, is $R_{\rm Alf, q}\simeq 21~R_*$. This implies that the radio emission at both 3.5 and 6 cm forms significantly closer than $20~R_*$ from the star's surface, if the origin of the radio emission is charged particles in the outflowing stellar wind.

The ultraviolet C~{\sc iv} $\lambda\lambda$1548, 1550 and  Si~{\sc iv} $\lambda\lambda$1394, 1402 UV resonance doublets are both sensitive indicators of the stellar wind and its structuring and modulation by the magnetic field \citep[e.g.]{Schnerr2008}. A search of the MAST archive indicates that HD 133880 was unfortunately not observed by the IUE satellite, nor with any other instrument with coverage in this spectral region. On the other hand, our optical spectra contain H$\alpha$ as well as a number of Paschen series H lines. An examination of H$\alpha$ reveals weak variability in the inner $\pm 500$~km/s of the profile.

\begin{figure}
\centering
\includegraphics[width=3.3in]{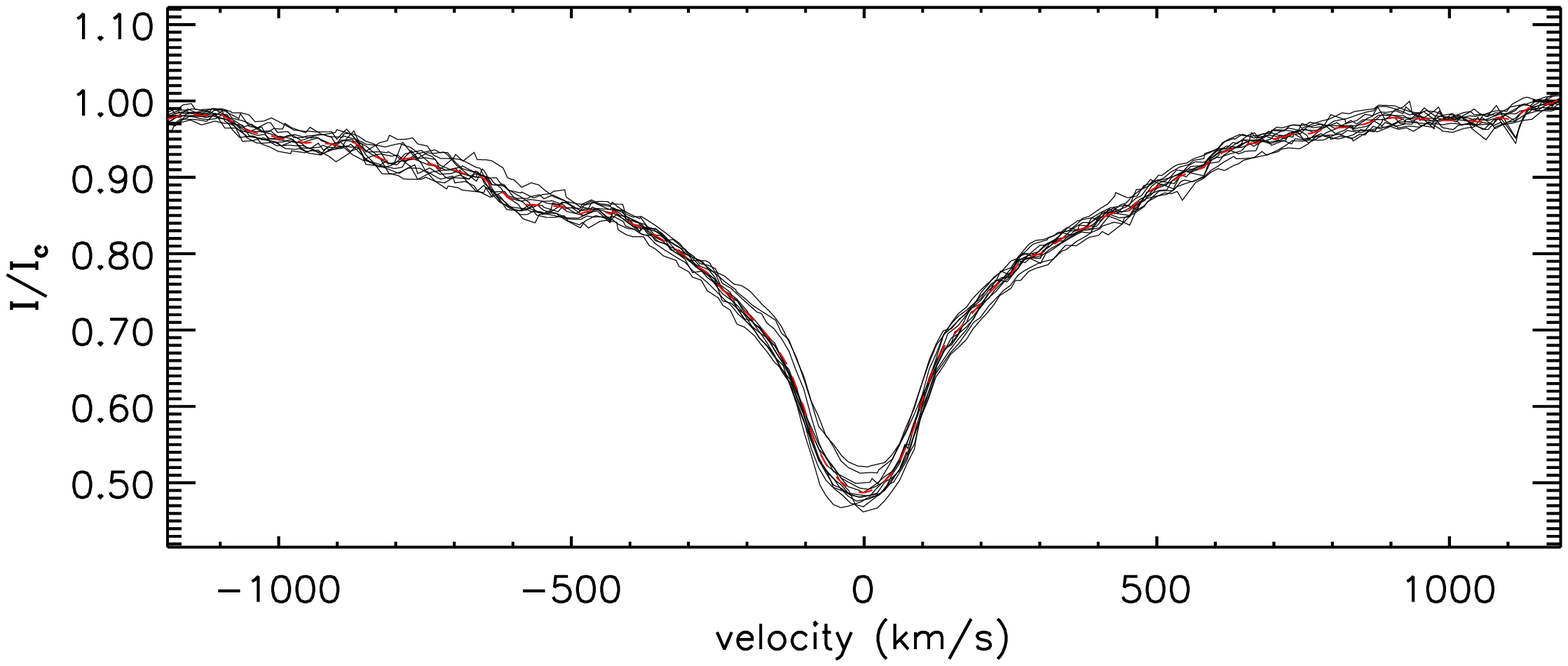}\\
\includegraphics[width=3.3in]{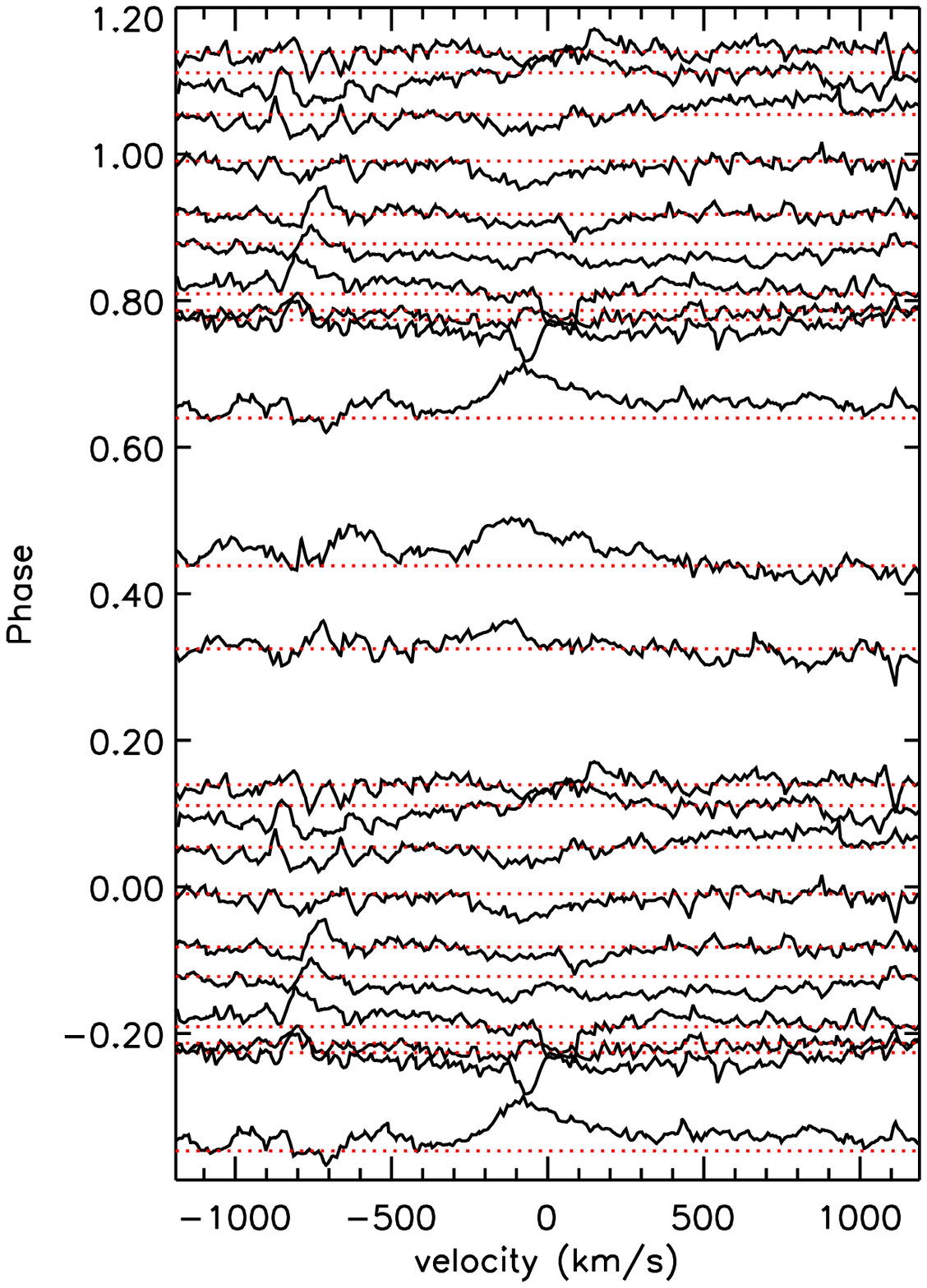}\\
\includegraphics[width=3.3in]{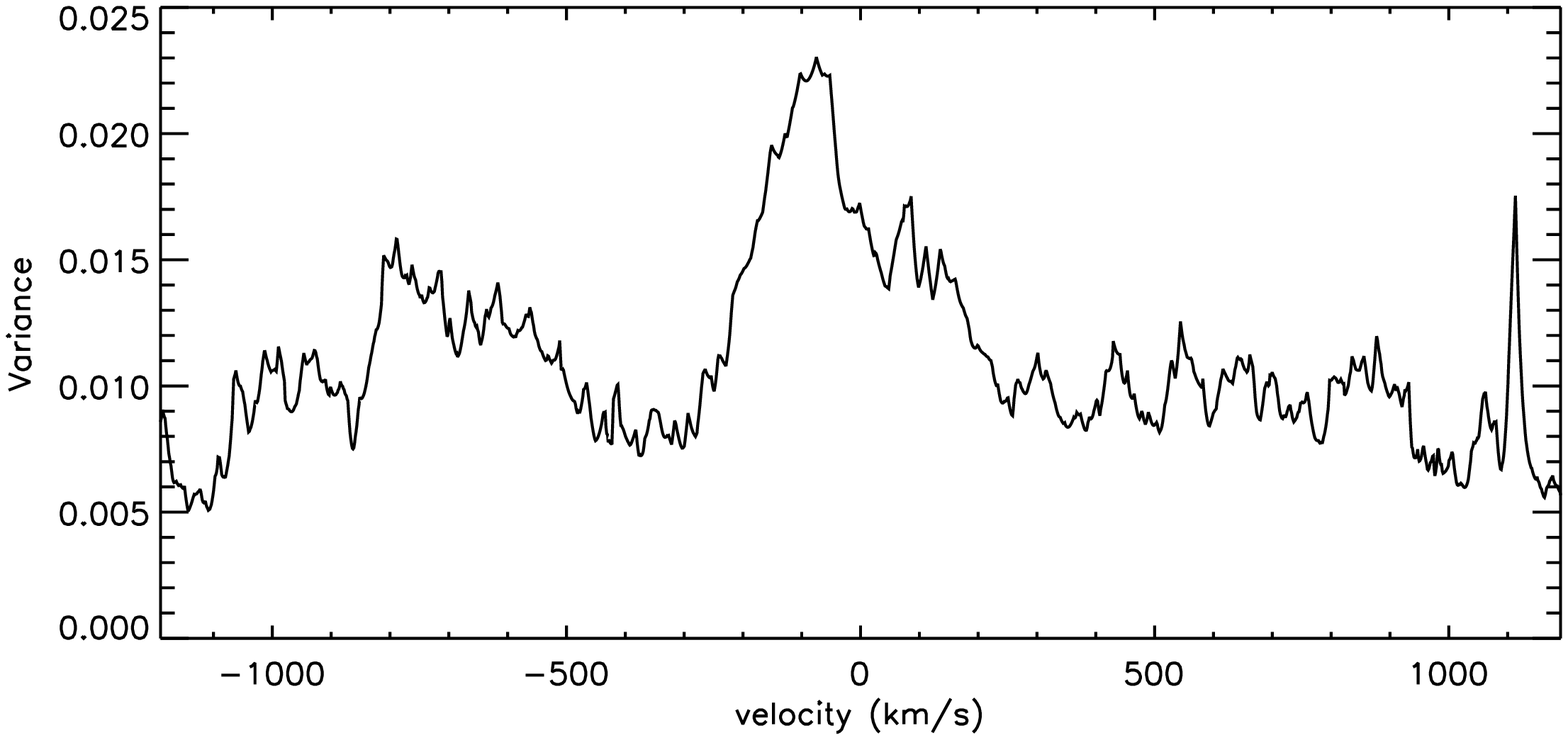}
\caption{{\bf Top:} Observed H$\alpha$ line profiles with the mean profile indicated as dashed red. {\bf Middle:} Shown are the continuum normalised spectra after subtracting the average of all the H$\alpha$ line profiles, displayed in such a way that the profiles are plotted at a height that corresponds to the phase of the observation. Also included is a dotted line corresponding to a zero difference to highlight the line profile variations. {\bf Bottom:}  Shown is the variance of the continuum normalised spectra from the mean profile.}
\label{halpha_dyn_plot}
\end{figure}

\begin{figure}
\centering
\includegraphics[width=3.3in]{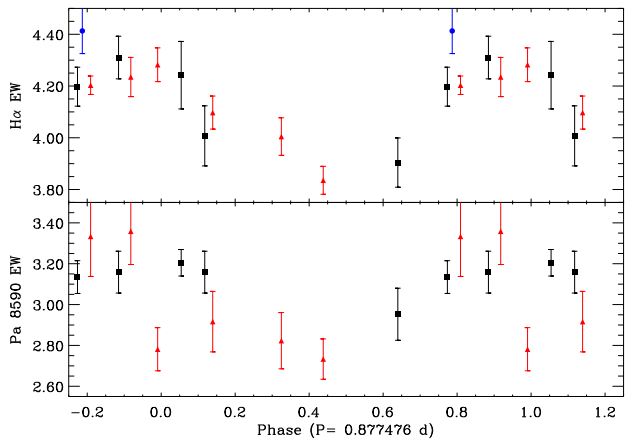}
\caption{Phased equivalent widths measurements of the core region of H$\alpha$ (top) and the total Paschen line at 8590\,\AA\ from the ESPaDOnS (black squares) and FEROS (red triangles) spectra.}
\label{paschen_eqw}
\end{figure}

In Fig.~\ref{halpha_dyn_plot} we illustrate the line profile variations that are observed in H$\alpha$. Fig.~\ref{paschen_eqw} shows the phased equivalent width of H$\alpha$, along with that measured from the Pa 8590\,\AA\ line. A telluric line removal algorithm was applied to each profile before analysis to reduce the variations caused by such features. While particular care has been taken to keep the normalisation of the ESPaDOnS and FEROS spectra as consistent as possible, it is difficult to judge whether the small fluctuations that are found in the far wings are real or an artefact of the normalisation. At phases 0.0 to 0.35 we find that the line cores show excess absorption relative to the average profiles, reaching maximum absorption at phase 0.0. Between phases 0.35 to 0.8, the core region of the H$\alpha$ profile shows emission relative to the average profile. Unfortunately, our phase sampling is not sufficient to distinguish if these features propagate through the profile, as would be expected from rigidly rotating magnetically confined plasma. 

The H$\alpha$ profile variability of HD 133880 might be a consequence of modifications of the atmospheric structure due to the strong chemical non-uniformities we have determined to exist in its atmosphere. However, we know that the disc-integrated photospheric He abundance is constrained to be below 10\% that of the Sun, i.e. less than 1\% of hydrogen by number, at all phases. So even a significant variation of the He abundance could have no important effect on the atmospheric structure. 

For the He star a Cen (which has a huge variation of He abundance over its surface and strongly variable metallic lines; \citet{Bohlender2010}) H$\alpha$ varies only in the core of the line (over a region perhaps 2 or 3 times the width of a typical metallic photospheric line) and the variations quite closely mimic the variations observed in the star's iron lines.  i.e. the core of H$\alpha$ gets stronger at the same time as Fe~{\sc ii} lines are strong.  This may be what we are seeing in HD 133880 as well, since the surface field and metal lines are weakest when H$\alpha$ is also weakest and most of the variations occur within $\pm 200$ km/s of line centre. Given the dominance of the quadrupolar field component, it is possible that the contribution of magnetospheric material to the overall variability of H$\alpha$ may be reduced somewhat, since material will be trapped in a more complex circumstellar geometry: likely four or more ``clouds'' rather than the two clouds seen in stars with dominant dipolar fields.  For example, the H$\alpha$ variability of HD~37776 (Grunhut et al., in preparation), a B-type star with a complex quadrupolar field, is quite modest compared to other stars with more dipolar fields (\citep[e.g. $\sigma$ Ori E;]{oksala2011}) despite the fact that it has similar physical properties to the other stars. Therefore, at present we simply conclude that the origin of the H line variations is unclear.

\section{Discussion}
This paper presents our continuing work in the study of chemical abundance evolution with time of magnetic Ap/Bp stars in open clusters.  The goal of this project is to model the magnetic field structure, to derive an estimate of the surface abundance distribution of a number of elements and to provide a preliminary analysis to serve as a foundation for further more detailed modelling.  In particular, more spectropolarimetric observations will allow for a more sophisticated interpretation of the surface abundance variations and magnetic field structure via Magnetic Doppler Imaging (MDI; \citet{pisk2002}).

HD~133880 is a very young, rapidly rotating Bp star (\vs\ $\simeq$ 103~\kms) that hosts a strong magnetic field. The magnetic field variations were modelled using a co-linear axisymmetric multipole expansion with $i$=55$^{\circ} \pm 10^{\circ}$, $\beta$=78$^{\circ} \pm 10^{\circ}$, $B_{d} = -9600 \pm 1000$~G, $B_{q} = -23200 \pm 1000$~G, and $B_{oct} = 1900 \pm 1000$~G.  We used Hipparcos photometric measurements together with our own \bz\ measurements, as well as photometric and radio emission data presented by \citet{waelkens1985} and \citet{Lim1996} respectively, to refine the period to $P = 0.877476 \pm 0.000009$~days.  The abundance distributions of He, O, Mg, Si, Ti, Cr, Fe, Ni, Pr, and Nd were modelled using three co-axial rings around the two magnetic poles and the magnetic equator.  The Fe-peak elements are overabundant compared to the solar ratios with strong variations on a global scale observed between hemispheres.

We have twelve Stokes $I$ spectra of HD~133880 spread well in phase that sample both magnetic poles and the magnetic equator.  Our adopted magnetic field geometry, although an approximation to the actual field structure of HD~133880, and our three co-axial ringed abundance model provides the most detailed abundance analysis possible given the limitations of the model.  Most elements studied appear to have a distinctly non-solar abundance in most rings.  The abundance of Mg is slightly above solar at both magnetic poles but near solar abundance at the magnetic equator. Upper limits to the abundances of He and Ni were found.  HD~133880 is clearly a He-weak star with an abundance at least a factor of 8 lower than the solar ratio.  Ni is no more than 10 times overabundant compared to the Sun.  All other elements studied (O, Si, Ti, Cr, Fe, Pr, Nd) are more abundant at the negative magnetic pole than the positive magnetic pole.  Most interesting are the variations observed in O abundance.  Most Ap/Bp stars show O abundances near or below the solar abundance ratios  \citep{oxygen1962,oxygen1990}.  However, a careful study of several O features reveal that oxygen is more abundant than solar at both magnetic poles (most notably at the negative magnetic pole where O is 0.7~dex overabundant) but a solar abundance near the magnetic equator.  This behaviour is certainly uncommon in Bp stars, but is unambiguously determined in this analysis.

In the context of the larger study of the time evolution of atmospheric abundances of Ap/Bp stars, it is interesting to compare to work done by \citet{Bailey2011} on the Ap star HD~318107 (=NGC 6405 77).  This star is also hot (\teff\ $\simeq$ 11800~K), young ($\log t \simeq 7.8$ (yrs)), of comparable mass ($2.95 \pm 0.15~M_{\odot}$), and host to a large complex magnetic field (\bz\ varies from about 1 to 5~kG).  In contrast to HD~133880, it is a slow rotator (\vs\ $\simeq$ 7~\kms) and only one magnetic hemisphere is observed.  The global properties of atmospheric abundances is similar between both stars: the mean abundances for Fe-peak (Ti, Cr, Fe) and rare-earth (Nd, Pr) elements are comparable within uncertainties and clearly overabundant compared to the solar ratios.  There are, however, notable differences in the abundances of O and Si between both stars.  For HD~318107, the O abundance is more typical of Ap stars, being about 0.3 dex lower than the solar value as compared to HD~133880 where O appears to be overabundant.  Si is about 1~dex more abundant in HD~133880 compared to HD~318107 which is understandable in that the former has the Si~$\lambda$4200 peculiarity whereas the latter does not.

\citet{folsom2007} studied the chemical abundances of stars  in an older cluster NGC~6475 ($\log t \simeq 8.47$~(yrs)) that have similar masses to HD~133880.   Mean abundances for multiple elements in the Ap stars HD~162576 (=NGC~6475~55; $3.0 \pm 0.4~M_{\odot}$ ), HD~162725 (=HR~6663; $3.3 \pm 0.5~M_{\odot}$) and HD~162305 (=NGC~6475~14; $2.7 \pm 0.3~M_{\odot}$) were derived.  As compared to HD~133880 and HD~318107, the abundances of Cr for all three stars are similar, being about 2~dex larger than in the Sun.  However, the abundances of Ti and Fe appear to be slightly lower and only 0.5~dex larger than the solar ratios, respectively.  This is in contrast to HD~133880 and HD~318107 where Ti and Fe are about 1.5 and 1~dex overabundant compared to the Sun, respectively.  A larger dataset is required before any definitive conclusions can be made regarding the time evolution of abundances in Ap stars.  

HD~133880 was previously modelled by \citet{Landstreet1990} who characterised the magnetic field variations of this star with measurements of the longitudinal magnetic field from H$\beta$.  Unlike most magnetic Ap/Bp stars whose magnetic field structure is predominantly dipolar, \citet{Landstreet1990} found a magnetic field dominated by the quadrupole component.  Despite the large \vs\ of the star, we were able to further characterise the magnetic field of HD~133880.  The higher SNR and spectral resolution of the ESPaDOnS spectra made it possible to compare spectral lines of the same element that have a large and small Land\'e $z$ factor to characterise the surface magnetic field modulus variations, \bs.  By finding a model that characterises the observed \bz\ and \bs\ variations, we are able to find the inclination, $i$, of the line of sight to the rotation axis as well as the tilt angle of the magnetic field axis, $\beta$, to the rotation axis.  This model is merely an approximation to the very complex field structure of HD~133880, but is sufficient to provide a meaningful first analysis to the atmospheric abundance distribution of elements.

Figure~\ref{magfield} shows clearly the qualitative nature of both the \bz\ and \bs\ variations.  There is a rapid change in polarity in the \bz\ curve with a steep transition from the positive to negative hemisphere.  The strong quadrupolar component of the magnetic field causes a distinct non-sinusoidal variation with the field structure flattening out near the positive magnetic pole.  The fact that there is a clear pole reversal indicates that the sum of the inclination and magnetic field axis must be greater than 90$^{\circ}$: $i + \beta > 90^{\circ}$, and that the line-of-sight goes well into both magnetic hemispheres.  The parameters of the adopted magnetic field geometry are presented in Sect. 6.2.                     
\begin{table}
\caption{Measurements of \bz\ using LSD masks composed of lines of single elements (Si, Ti, Cr, and Fe) from the polarised ESPaDOnS and HARPS spectra (cf Table~\ref{fulllsd}).}
\scriptsize
\begin{tabular}{ccccc}
\hline\hline
Phase               &   \bz(Si)   &   \bz(Ti)  &   \bz(Cr)   &   \bz(Fe)   \\
  &     (G)     &     (G)      &      (G)      &   (G)        \\
\hline
0.111  & \multicolumn{1}{r}{$-3419 \pm 76$} & \multicolumn{1}{r}{$-5724 \pm 535$} & \multicolumn{1}{r}{$-4773 \pm 229$} & \multicolumn{1}{r}{$-3498 \pm 50$} \\
0.640  & \multicolumn{1}{r}{$1790 \pm 66$} & \multicolumn{1}{r}{$2255 \pm 304$} & \multicolumn{1}{r}{$2653 \pm 196$} & \multicolumn{1}{r}{$1993 \pm 44$} \\
0.787  & \multicolumn{1}{r}{$1020 \pm 93$} & \multicolumn{1}{r}{$-506 \pm 308$} & \multicolumn{1}{r}{$1026 \pm 208$} & \multicolumn{1}{r}{$975 \pm 78$} \\
0.878  & \multicolumn{1}{r}{$-1776 \pm 63$} & \multicolumn{1}{r}{$-2883 \pm 319$} & \multicolumn{1}{r}{$-2409 \pm 175$} & \multicolumn{1}{r}{$-2010 \pm 45$} \\
\hline\hline
\label{Bz-per-el}
\end{tabular}
\normalsize
\end{table}
\bz\ measurements using all metallic lines (see Table~\ref{fulllsd}) agree well with the measurements by \citet{Landstreet1990} using H$\beta$, suggesting that they sample the magnetic field in similar ways (see Fig.~\ref{magfield}).  Longitudinal field \bz\ measurements of spectral lines of only a single element provide clues to the inhomogeneous structure of the magnetic field and the surface distribution of elements.  Using all four of our polarimetric observations, we have remeasured the longitudinal field \bz\ using spectral lines of Si, Ti, Cr, and Fe.  The results are listed in Table~\ref{Bz-per-el}.

It is clear from Table~\ref{Bz-per-el} that all four elements sample the field in different ways.  The differences in measured field strengths can be attributed to the complex magnetic field (as reported by \citet{Landstreet1990} and discussed in this paper) and the inhomogeneous surface distribution of elements which is a common feature of Bp stars.  Near the positive magnetic pole (phase 0.537) at phase 0.640, as well as phase 0.878, the measured \bz\ values for all four elements differ significantly.  In the negative magnetic hemisphere (near phase 0.037) Si and Fe sample the field in similar ways, whereas Cr and Ti sample the magnetic field differently with \bz\ measurements approximately 1 and 2~kG stronger than Si and Fe.  Along the sharp decline from the positive to negative magnetic pole (near phase 0.787; see Figure~\ref{magfield}), it appears that only Ti differs greatly showing a polarity reversal as compared to the other elements at the same phase.             

The measurements of the longitudinal field using different elements reveal that particular elements at different phases sample the field in very distinctive ways which suggests a truly complex magnetic field that cannot be accurately modelled by a simple axisymmetric magnetic field (such as the one adopted in this paper and by \citet{Landstreet1990}).  It also strongly suggests an inhomogeneous distribution of various elements on the stellar surface that substantially affect the magnetic field measurements.  We also compared the computed to the observed Stokes $V$ profiles for the three polarimetric ESPaDOnS spectra for multiple lines (Mg, Si, Ti, Cr, and Fe).  The agreement is poor, with the computed Stokes $V$ profiles being much stronger than the observed, further suggesting that our adopted magnetic field model is only a rough approximation.  This result is not unique.  A recent paper by \citet{hd37776-2011} on HD~37776 (=HIP~26742) reveal that the complex multipole model derived from the longitudinal field predicts Stokes $V$ signatures that are much too strong compared to the observed Stokes $V$ profiles.  In this case, the multipole magnetic field model (similar in structure to the one used for HD~133880) predicts a large surface field, whereas the MDI results produce a much more modest surface field.  This is also seen, although less severely, for $\sigma$ Ori E \citep{oksala2011}.  For this star, detailed analysis using MDI produced a magnetic field model that closely resembled the observed Stokes $V$ profiles.  The acquisition of new, high SNR polarimetric observations of HD~133880 are necessary to characterise unambiguously the surface abundance variations and magnetic field structure using MDI.   

The very strong field and rapid rotation of HD~133880 produce interesting magnetospheric characteristics.  In our preliminary analysis, we were not able to conclude that the optical spectra contain detectable information about the magnetosphere.  However, HD~133880 is one of only a few Bp stars for which radio flux variations have been noted.  There are small but distinct differences between the 3~cm and 6~cm radio emission, such as the systematically lower 3 cm than 6 cm fluxes at phases 0.0 and 0.5.  The 3~cm emission suggests a ``bump'' at about phase 0.6 and the peak at about phase 0.75 is narrower at 3~cm than at 6~cm.  The fact that the radio spectrum is flat between 6 and 3 cm suggests that the emission is optically thick, and therefore we are observing the magnetosphere at two different radio surfaces rather than an integration throughout the entire volume.  These subtle differences in the radio light curves at 6 and 3 cm then imply that these two surfaces cannot have exactly the same shapes and that, presumably, the magnetic field is not perfectly axisymmetric.  HD~133880 is a very exciting target for more detailed magnetospheric modelling.
 
\section*{Acknowledgments}   
JDB, JDL, and GAW are grateful for support by the Natural Sciences and Engineering Research Council of Canada.  We thank V\'eronique Petit, for communicating calculations of the magnetospheric parameters for HD 133880 prior to publication.  We thank the HARPSpol team for generously providing a spectrum for HD 133880 used in this analysis.  We thank the referee, Dr. Evelyne Alecian, for her useful comments and suggestions.  

\label{lastpage}

\bibliographystyle{mn2e}
\bibliography{hd133880}

\end{document}